\documentclass{article}

\usepackage[top=1in, bottom=1in, left=1in, right=1in]{geometry}
\usepackage{setspace}
\usepackage{amsmath}
\usepackage{accents}
\usepackage{amsfonts}
\usepackage{amsthm} % for proof environment
\usepackage{amssymb} % for \subsetneq
\usepackage{epsfig}
\usepackage{url}
\usepackage{bm}
\usepackage{bbm}
\usepackage{xcolor}

\usepackage{tikz-cd}
\tikzcdset{ampersand replacement=\&}
\usepackage{graphicx}
\usepackage{subcaption}
\usepackage[ruled,vlined]{algorithm2e}
\usepackage{svg}

\usepackage[colorlinks,pagebackref=true]{hyperref}

\newcommand{\e}{{\rm e}}
\newcommand{\E}{{\mathbb E}}
\newcommand{\Pa}{{\mathbb P}}
\newcommand{\Q}{{\mathbb Q}}

\newcommand{\R}{{\mathbb R}}
\newcommand{\N}{{\mathbb N}}

\newcommand{\Bcal}{{\mathcal B}}

\newcommand{\Ecal}{{\mathcal E}}
\newcommand{\Fcal}{{\mathcal F}}
\newcommand{\Gcal}{{\mathcal G}}

\newcommand{\Ncal}{{\mathcal N}}

\newcommand{\Pcal}{{\mathcal P}}

\newcommand{\Scal}{{\mathcal S}}
\newcommand{\Tcal}{{\mathcal T}}

\newcommand{\1}{{\mathbbm 1}}

\usepackage{mathtools}

\usepackage{footmisc}
\newtheorem{proposition}{Proposition}[section]

\newtheorem{exampleemph}[proposition]{Example}   % upshape
\begin{document}
\title{Ensemble learning for portfolio valuation and risk management }
\author{ Lotfi Boudabsa\footnote{EPFL. Email: lotfi.boudabsa@epfl.ch} \and Damir Filipovi\'c\footnote{EPFL and Swiss Finance Institute. Email: damir.filipovic@epfl.ch}}
\date{12 April 2022}
\maketitle

\begin{abstract}
We introduce an ensemble learning method for dynamic portfolio valuation and risk management building on regression trees. We learn the dynamic value process of a derivative portfolio from a finite sample of its cumulative cash flow. The estimator is given in closed form. The method is fast and accurate, and scales well with sample size and path space dimension. The method can also be applied to Bermudan style options. Numerical experiments show good results in moderate dimension problems. 
\end{abstract}

\noindent {\bf Keywords:} dynamic portfolio valuation, ensemble learning, gradient boosting, random forest, regression trees, risk management, Bermudan options

% \noindent {\bf MSC (2010) Classification:} 68T05, 91B28\\

% \noindent {\bf JEL Classification:} C63, G12

\section{Introduction}\label{section_introduction}
The valuation and risk management of derivative portfolios form a challenging task in banks, insurance companies, and other financial institutions. The issues are formally explained as follows. Most economic scenario generators can be represented as stochastic models with finitely many time periods $t=0,1,\dots,T$, where randomness is generated by some underlying stochastic driver $X=(X_1,\dots,X_T)$. The components $X_t$ are mutually independent, but not necessarily identically distributed, taking values in $\R^{d}$, for some $d\in\N$.\footnote{We endow $\R^d$ with the Borel $\sigma$-algebra $\Bcal(\R^d)$.} We assume that $X$ is realized on the path space $\R^{d\times T}$ such that $X_t(x)=x_t$ for a generic sample point $x=(x_1,\dots,x_T)$. We denote the distribution of $X$ by $\Q = \Q_1\times\dots\times \Q_T$, and we assume that $\Q$ represents the risk-neutral pricing measure with respect to some fixed numeraire, such as the money market account. All financial values and cash flows henceforth are discounted by this numeraire, if not otherwise stated. The stochastic driver $X$ generates the filtration $\Fcal_t=\Bcal(\R^{d})^{\otimes t}$ which represents the flow of information.\footnote{Henceforth, we let $\Fcal_0$ denote the trivial $\sigma$-algebra. However, we could easily extend the setup to include randomness at $t=0$, by setting $X=(X_0,X_1,\dots,X_T)$. Here $X_0$ could include cashflow specific values that parametrize the cumulative cashflow function $f(X)$, such as the strike price of an embedded option or the initial values of underlying financial instruments. We could then sample $X_0$ from a Bayesian prior $\Q_0$.}

We consider a portfolio whose cumulative cash flow is modeled by some measurable function $f:\R^{d\times T}\to\R$ such that $f\in L^2_\Q$. Its dynamic value process $V$ is then given by the martingale
\begin{equation}\label{eqcondY_EU}
\textstyle V_t=\E_\Q[f(X) \mid\Fcal_t] , \quad t=0,\dots,T.
\end{equation}
Computing $V$ is challenging, because the conditional expectations in \eqref{eqcondY_EU} are not available in closed form in general. This is the case for most exotic and path-dependent options, such as barrier reverse convertibles, or the following max-call.

For illustration, let us consider the multivariate Black--Scholes model, where $X_t$ are i.i.d.\ standard normal on $\R^d$. There are $d$ nominal stock prices given by
\begin{equation}\label{bs_model}
\textstyle
    S_{i,t}=S_{i,t-1}\exp[ \sigma_i^\top X_t \sqrt{\Delta_t} + (r- \|\sigma_i\|^2 /2)\Delta_t],\quad t=1, \dots, T,
\end{equation}
for some initial values $S_{i,0}$, volatility vectors $\sigma_i\in\R^d$, $i=1,\dots,d$, constant risk-free rate $r$, and time step size in units of a year $(\Delta_1, \dots, \Delta_T)$. Then there exists no closed-form expression for the value process $V$ of the max-call option whose discounted payoff at $T$ is
\begin{equation}\label{intro_example}
\textstyle f(X) =\e^{-r\sum_{t=1}^T \Delta_t}(\max_{i} S_{i, T} - K)^+,
\end{equation}
for some strike price $K$.

We solve this issue via a novel method to learn the portfolio value process $V$. First, we use ensemble estimators with regression trees to learn the function $f$ from a finite sample $\bm X=(X^{(1)},\dots,X^{(n)})$, drawn from $\Q$, along with the corresponding function values $\bm f = (f(X^{(1)}), \dots, f(X^{(n)}))$.\footnote{More precisely, $\bm X$ consists of i.i.d.\ $\R^{d\times T}$-valued random variables $X^{(i)}\sim \Q$ defined on the product probability space $(\bm E , \bm\Ecal,  {\bm  Q})$ with $ \bm E = \R^{d\times T} \otimes \R^{d\times T} \otimes\cdots $, $ \bm \Ecal =\Bcal(\R^{d})^{\otimes  T} \otimes\Bcal(\R^d)^{\otimes  T}\otimes\cdots $, and $  {\bm  Q} =  \Q \otimes\Q\otimes\cdots $. \label{footnote_1}} Here we consider the two most popular ensemble estimators, namely Random Forest defined in \cite{bre_2001}, and Gradient Boosting defined in \cite{fri_2001}. We denote these estimators of $f$ by $f_{\bm X}$. In either case, the expression of $f_{\bm X}$ is of the form
\begin{equation}\label{generic_el_estimator}
\textstyle    f_{\bm X} = \sum_{i=1}^{N} \beta_i \1_{\bm A_i},
\end{equation}
where $\beta_i$ are real coefficients, and $\bm A_i$ are hyperrectangles of $\R^{d\times T}$ that cover $\R^{d\times T}$, but are not necessarily disjoint. See \eqref{box_notation} for the definition of a hyperrectangle. Second, we use $f_{\bm X}$ to define the process $V_{\bm X}$ as follows,
\begin{equation}\label{eqYhat}
\textstyle
  V_{\bm X,t}=\E_\Q[f_{\bm X}(X) \mid\Fcal_t ],\quad t=0,\dots, T.
\end{equation}
The process $V_{\bm{X}}$ in \eqref{eqYhat} is an estimator of the value process $V$ in \eqref{eqcondY_EU}. This estimator is fast to construct for two reasons. First, the training of $f_{\bm{X}}$ is fast, which is due to the use of readily accessible and highly optimized implementations of Random Forest and Gradient Boosting. Specifically, for Random Forest we use the RandomForestRegressor class in scikit-learn \cite{scikit_learn}, and for Gradient Boosting we use the XGBRegressor class in XGBoost (eXtreme Gradient Boosting) \cite{che_gue_2016}. Second, the conditional expectations in \eqref{eqYhat} are given in closed form, in the sense that they can be efficiently evaluated at very low
computational cost, see Section \ref{section_closed_form_Veu_Vus} for details. Furthermore, there is empirical evidence showing that $f_{\bm X}$ is an accurate estimator of $f$, i.e., $f_{\bm{ X}}$ achieves a small $L^2_\Q$-error $\|f-f_{\bm{ X}}\|_{2, \Q}$, in many problems arising from different fields (scientific fields, and machine learning and data mining challenges). See, e.g., \cite{bia_sco_2016} and references therein for Random Forest, and \cite{che_gue_2016} for Gradient Boosting. This implies that $V_{\bm{X}}$ is an accurate estimator of $V$. Indeed, thanks to Doob's maximal inequality, see, e.g., \cite[Corollary II.1.6]{rev_yor_94}, the path-wise maximum $L^2_\Q$-error is bounded by
\begin{equation}\label{doobineq}
\textstyle
 \|\max_{t=0,\dots,T} | V_t-V_{\bm{X},t}|\|_{2,\Q}\le   2\| f -f_{\bm{X}} \|_{2,\Q}.
\end{equation}

There are many risk management tasks building on the dynamic value process $V$. In this paper, we focus on risk measurement as a generic example.\footnote{Another important task in portfolio risk management is hedging, which is sketched in more detail in our previous paper \cite{bou_fil_22}.} For two dates $t_0<t_1$, we denote by $\Delta V_{t_0,t_1}  = V_{t_1}-V_{t_0}$ the gain from holding the portfolio over the period $[t_0,t_1]$. Portfolio risk managers and financial market regulators alike quantify the risk of the portfolio $V$ over $[t_0, t_1]$ by means of an $\Fcal_{t_0}$-conditional risk measure, such as value at risk or expected shortfall, evaluated at $\Delta V_{t_0,t_1}$.\footnote{For the definition of value at risk and expected shortfall (also called conditional value at risk or average value at risk), we refer to \cite[Section~4.4]{foe_sch_04}, and Section \ref{secexamples} below.} In practice these risk measures are applied under the equivalent real-world measure $\Pa\sim\Q$. Using the Cauchy--Schwarz inequality and \eqref{doobineq}, we obtain $\|\max_{t=0,\dots,T} | V_t-V_{\bm X,t}|\|_{1,\Pa}\le \| \frac{d\Pa}{d\Q}\|_{2,\Q}  \|\max_{t=0,\dots,T} | V_t-V_{\bm X,t}|\|_{2,\Q}\le2\| \frac{d\Pa}{d\Q}\|_{2,\Q}\|f-f_{\bm X}\|_{2,\Q}$, so that $V_{\bm X}$ and $V$ are close in $L^1_\Pa$ as soon as $f_{\bm X}$ and $f$ are close in $L^2_\Q$. Hence risk measures that are continuous with respect to the $L^1_\Pa$-norm, such as value at risk (under mild technical conditions) and expected shortfall, see, e.g., \cite[Section 6]{cam_fil_17}, return similar values when applied to $V_{\bm X}$ instead of $V$.

Related literature on portfolio risk measurement includes \cite{bro_du_moa_15} who introduce a regression-based nested Monte Carlo simulation method for the estimation of the unconditional expectation of a Lipschitz continuous function $f(L)$ of the 1-year loss $L=-\Delta V_{0,1}$. They also provide a comprehensive literature overview of nested simulation problems, including \cite{gor_jun_10} who improve the speed of convergence of the standard nested simulation method using the jackknife method. Our method is different as it learns the entire value process $V$ in one go, as opposed to any method relying on nested Monte Carlo simulation, which estimates $V_t$ for one fixed $t$ at a time. Our method shares some similarities with the kernel-based method in our previous paper \cite{bou_fil_22}. There we applied kernel ridge regression to derive a closed-form estimator of the value process $V$. That kernel-based estimator satisfies asymptotic consistency and finite sample guarantees. However, due to cubic training time complexity, it cannot be applied to high dimensional problems, i.e., problems where the sample size $n$ or the path space dimension $d\times T$ are very large. The ensemble estimators in the present paper scale better. Our method also share similarities with the GPR-EI (Gaussian Process Regression-Exact Integration) method in \cite{gou_et_al_2020}, which gives a closed-form estimator of the entire value process $V$ of an American option under the Black--Scholes and Rough--Bergomi models.

Here and throughout we use the following conventions and notation. For any $p\in [1,\infty)$ and measurable function $f:\R^{d\times T}\to\R$, we denote $ \|f\|_{p,\Q}=
(\int_{\R^{d\times T}} |f(x)|^p\Q(dx))^{1/p}$. We denote by $L^p_\Q$ the space of \emph{$\Q$-equivalence classes} of measurable functions $f:\R^{d\times T}\to\R$ with $\|f\|_{p,\Q}<\infty$. If not otherwise stated, we will use the same symbol, e.g., $f$, for a function and its equivalence class. Let $ a=(a_1,\dots, a_T)$ and $ b=(b_1, \dots, b_T)$, where $a_t,b_t\in\overline{\R^d}$, so that $ a, b\in \overline{\R^{d\times T}}$. Assume that $a_t<b_t$, i.e., $a_{j,t}<b_{j,t}$ for every $j=1,\dots, d$, for every $t=1,\dots, T$. A hyperrectangle $\bm A=( a,  b]$ of $\R^{d\times T}$ is a subset of $\R^{d\times T}$ of the form

\begin{equation}\label{box_notation}
\textstyle
\bm A= \{(x_1, \dots, x_T)\in \R^{d\times T}\mid a_t < x_t \le b_t,\, t=1,\dots, T\}.
\end{equation}
It is convenient to write $\bm A = \prod_{t=1}
^T ( a_t,b_t]$, where $( a_t, b_t] = \prod_{j=1}^d ( a_{j, t}, b_{j, t}]$ is a subset of $\R^{d}$.\footnote{If $b_{j,t}=\infty$, then $(a_{j,t}, b_{j,t}]$ is defined as $(a_{j,t}, \infty)$.}

The remainder of the paper is as follows. Section~\ref{section_new_estimators} presents the ensemble estimators we use to learn the function $f$. Section~\ref{section_closed_form_Veu_Vus} shows that the value process estimator $V_{\bm X}$ is in closed form for two large classes of financial models. Section~\ref{secexamples} provides numerical examples for the valuation of exotic and path-dependent options in the multivariate Black--Scholes model. Section \ref{sec_outlook} discusses future research directions. Section~\ref{secconc} concludes. Appendix~\ref{sec_regnow_reglat} compares our method to its regress-now variant. And Appendix \ref{sec_optimal_stopping} shows how our method can be applied to Bermudan options.

%%%%%% what is left to review

\section{Ensemble estimators based on regression trees}\label{section_new_estimators}

Following up on Section \ref{section_introduction}, we let $f\in L^2_\Q$. We now present the construction of the estimator $f_{\bm X}$ in \eqref{generic_el_estimator}, which is used to define the value process estimator $V_{\bm X}$ in \eqref{eqYhat}. As mentioned above, $f_{\bm X}$ is either a Random Forest or a Gradient Boosting. Random Forest is defined in \cite{bre_2001} using CART regression trees (CART stands for Classification And Regression Trees). The CART method is defined in \cite{bre_et_al_1984}. Gradient Boosting is defined in \cite{fri_2001} using ``a small regression tree, such as those produced by CART''. In order to make the paper self-contained, we first recap the CART method.\footnote{According to the survey \cite{loh_2014}, the first regression tree method is called AID (Automatic Interaction Detector) and was defined in \cite{mor_son_1963}. However, it is \cite{bre_et_al_1984} that has been the most influential in what we now call the regression tree literature. In this literature there are many different methods that produce regression trees. To the best of our knowledge, today the two most popular regression tree methods are CART defined in \cite{bre_et_al_1984}, and C4.5 defined in \cite{qui_1993}.} We then recap the construction of a Random Forest and Gradient Boosting. Throughout we assume as given a finite i.i.d.\ sample $\bm X = (X^{(1)}, \dots, X^{(n)})$ drawn from $\Q$, along with the function values $\bm f = (f(X^{(1)}), \dots, f(X^{(n)}))$. We denote the corresponding empirical distribution by $\Q_{\bm X}=\frac{1}{n}\sum_{i=1}^n \delta_{X^{(i)}}$.

\subsection{CART regression tree} 
The CART method gives a piece-wise constant function that has the following form,
\begin{equation}\label{reg_tree}
\textstyle
    f_{\bm X,  \Pi} = \sum_{\bm A \in  \Pi} \E_{\Q_{\bm X}}[f(X^{(0)}) \mid X^{(0)} \in \bm A] \1_{\bm A},
\end{equation}
where $ \Pi$ is a finite hyperrectangle partition of $\R^{d\times T}$, $X^{(0)}$ has distribution $\Q_{\bm X}$, and we have $\E_{\Q_{\bm X}}[f(X^{(0)}) \mid X^{(0)} \in \bm A] = \E_{\Q_{\bm X}}[f(X^{(0)}) \1_{\bm A} (X^{(0)})]/\Q_{\bm X}[\bm A]$ if $\Q_{\bm X}[\bm A]>0$, and we use the convention $\E_{\Q_{\bm X}}[f(X^{(0)}) \mid X^{(0)} \in \bm A] = 0$ if no point $X^{(i)}$ lies in $\bm A$. The partition $ \Pi$ is constructed recursively by refining the trivial partition $ \Pi_0 = \{\R^{d\times T}\}$ in the following way. Assume at step $t\ge0$ the partition $ \Pi_{t}$ is of size $K_t\ge 1$. Then pick $\bm A \in \Pi_t$ and perform an axis-aligned split, denoted by $(j, s, z)$, to obtain the two hyperrectangles $\bm A_{L} = \{x \mid x \in \bm A,\, x_{j, s} \le z\}$ and $\bm A_{R} = \{x \mid x \in \bm A,\, x_{j,s} > z\}$. This gives a new partition $\Pi_{t+1} = (\Pi_t \setminus \bm A) \cup \{\bm A_{L}, \bm A_{R}\}$ of size $K_{t+1} = K _t + 1$. To find the optimal split $(j,s, z)$ of $\bm A$, one minimizes the within-group variance
\begin{equation}\label{fct_to_max}
    \begin{aligned}
    (j,s,z)\mapsto V_{\bm A}(j,s, z) &=\E_{\Q_{\bm X}}[\1_{\bm A_{L}}(X^{(0)})(f(X^{(0)}) -  \E_{\Q_{\bm X}}[f(X^{(0)}) \mid X^{(0)} \in \bm A_L])^2] \\
    &\quad + \E_{\Q_{\bm X}}[\1_{\bm A_{R}}(X^{(0)}) (f(X^{(0)}) - \E_{\Q_{\bm X}}[f(X^{(0)}) \mid X^{(0)} \in \bm A_R])^2]   
\end{aligned}
\end{equation}
over $\Scal  = \{(j,s, z) \mid j=1, \dots, d,\, s=1, \dots, T,\, z\in\R\}$.

When to stop refining the partition $\Pi_t$?\footnote{Other regression tree methods construct the partition $\Pi_t$ relying on other minimization criteria than the within-group variance in \eqref{fct_to_max}. For instance, C4.5 \cite{qui_1993} relies on the gain ratio.} In practice, standard stopping rules include to not split a hyperrectangle $\bm A$ if it contains less than a certain number \textbf{nodesize} of points. Another rule is to stop refining $\Pi_t$ when it reaches a certain size $K$. Then, according to \cite{bre_et_al_1984}, $\Pi_t$ should be pruned by looking for an optimal sub-partition $\Pi_{t'}\subseteq \Pi_{t}$, $t'\le t$, so that $f_{\bm X, \Pi}$ in \eqref{reg_tree} is defined with $\Pi = \Pi_{t'}$. However we omit this step and define $f_{\bm X, \Pi}$ with $\Pi=\Pi_t$. In fact, as discussed in \cite[Section~4]{bre_2001} and \cite[Section~8]{fri_et_al_2002}, a CART regression tree should not be pruned when it is used to define a Random Forest or Gradient Boosting.

The CART regression tree is known for its interpretability, and ability to perform dimensionality reduction and handle outliers. However this estimator is very sensitive to the sample $\bm X$: small perturbations of the sample $\bm X$ can lead to large changes in $f_{\bm X, \Pi}$. In response to this issue, Bagging (from bootstrap and aggregating) has been introduced in \cite{bre_1996}. Bagging is the aggregation, i.e., the average, of $M$ CART regression trees. The $m$-th tree is constructed using a sample $\bm{X}_m =(X^{(1)}_m, \dots, X^{(n)}_m)$, obtained by bootstrapping from $\bm X$, and the corresponding function values $ \bm f_m = (f(X^{(1)}_m), \dots, f(X^{(n)}_m))$. Bagging gives significantly better results than a single CART regression tree. Five years later, Random Forest was introduced in \cite{bre_2001}. Random Forest is an enhancement of Bagging. This will be our first ensemble estimator.

\subsection{Random Forest}\label{sec_rf}
\cite{bre_2001} defines a class of estimators called Random Forest. The same paper gives an example of Random Forest termed Random Forest-RI, where RI stands for Random Inputs. As highlighted in the survey \cite{gen_pog_2016}, nowadays the name Random Forest very often refers to Random Forest-RI. Therefore, we will call Random Forest-RI simply Random Forest. This estimator is the aggregation of $M$ regression trees, which are grown slightly differently than CART. Below we detail its construction.

Fix $\widetilde{n} \le n$ and let $(\bm{X}_1, \dots, \bm{X}_M)$ be $M$ samples, where each sample $\bm{X}_m =(X^{(1)}_m, \dots, X^{(\widetilde{n})}_m)$ is constructed by resampling $\widetilde{n} $ points from $\bm X$. The resampling can be with or without replacement. The resampling is called bootstrapping when it is done with replacement and $\widetilde{n}=n$, otherwise it is called subsampling (with or without replacement). Fix $p \in \{1, \dots,  d\times  T\}$, and grow $M$ regression trees, where the $m$-th tree $f_{\bm{X}_m, \Pi_m}$ is constructed as follows. Instead of $\bm X$ and $\bm f$, use the sample $\bm X_m$ and the corresponding function values $ \bm f_m = (f(X^{(1)}_m), \dots, f(X^{(\widetilde{n})}_m))$. And for every hyperrectangle $\bm A$ to split, draw uniformly $p$ coordinates $(j_1, s_1), \dots, (j_p, s_p)$ from $\{1, \dots, d\}\times\{1, \dots, T\}$, and minimize $(j,s,z)\mapsto V_{\bm A}(j,s,z)$ in \eqref{fct_to_max} over $\{(j_i, s_i,z)\mid i=1, \dots, p,\, z\in \R\}$ instead of $\Scal$. Then Random Forest, denoted by $f_{\bm{X}, \bm \Pi}$ with $\bm \Pi = ({\Pi}_1, \dots, {\Pi}_M)$, is the aggregation of the $M$ regression trees $f_{\bm{X}_m, \Pi_m}$,
\begin{equation}\label{rand_forest}
\textstyle
    f_{\bm X, \bm \Pi}  = \frac{1}{M} \sum_{m=1}^M f_{\bm{X}_m, {\Pi}_m}.
\end{equation}
In the case where $p= d\times T$ and the sampling scheme is bootstrapping, Random Forest is just Bagging. 
\subsection{Gradient Boosting}\label{sec_gb}

% According to Hastie et al. \cite{fri_et_al_2009} ``Boosting is one of the most powerful learning ideas introduced in the last
% twenty years". 

The idea of Boosting goes back to a theoretical question posed in \cite{kea_1988} and \cite{kea_val_1994}, called the ``Hypothesis Boosting Problem''. In the context of binary classification problems, the authors asked whether there exist a process able to turn a weak learner into a strong one. Such a process would be called Boosting. Here a weak learner is a classifier that performs only slightly better than random guessing. And a strong learner is a classifier that achieves a nearly perfect classification. A positive answer to this question was given in \cite{Sch_1990}. However the algorithm in \cite{Sch_1990} could not be implemented in practice, and it is AdaBoost, the algorithm defined in \cite{fre_sch_1996}, that is usually considered as the first workable Boosting algorithm. The success of AdaBoost with classification trees was such that Breiman called it ``best off-the-shelf classifier in the world", see \cite{fri_et_al_2002}. In order to better understand the performance of AdaBoost, a lot of research has been done. This includes the statistical framework developed in \cite{fri_et_al_2002}, which was further developed in \cite{fri_2001} to cover both classification and regression problems. In the later paper, Gradient Boosting is defined. This estimator is based on CART regression trees, and is constructed recursively, for $t\ge 1$, as follows,
\begin{equation}\label{gradient_descent}
\textstyle
  \begin{cases}
  f_{\bm X, t}(x) &= f_{\bm X, t-1}(x) - \gamma_t g_t(x), \quad x \in \R^{d\times T},\\
  \gamma_t &\in \arg\min_{\gamma\in \R_+}\E_{\Q_{\bm X}}\left[\psi(f(X^{(0)}),f_{\bm X, t-1}(X^{(0)}) - \gamma g_t(X^{(0)}))\right],
  \end{cases}
\end{equation}
where $f_{\bm X, 0}=\frac{1}{n}\sum_{i=1}^nf(X^{(i)})$, and $\psi :\R^2 \mapsto \R$, $(x,y)\mapsto \psi(x,y)$ is a given loss function. The function $g_t$ is a CART regression tree that estimates the function $x\mapsto \partial_y\psi(f(x), f_{\bm X, t-1}(x))$ using the sample $\bm X$, along with the function values $(\partial_y\psi(f(X^{(1)}), f_{\bm X, t-1}(X^{(1)})), \dots, \partial_y\psi(f(X^{(n)}), f_{\bm X, t-1}(X^{(n)})))$.\footnote{For the sake of brevity, we write $\partial_y\psi(f(x), f_{\bm X, t-1}(x))$ instead of $\frac{\partial \psi(\cdot, \cdot)}{\partial y}\mid_{(f(x), f_{\bm X, t-1}(x))}$.} And $\gamma_t$ is called the optimal step-size.

Often in regression problems, one picks the squared error loss function $ \psi(x, y)=\frac{1}{2}(x-y)^2$, which is what we do in Section \ref{secexamples}. In this case, $x\mapsto \partial_y\psi\big(f(x), f_{\bm X, t-1}(x)\big) = f_{\bm X, t-1}(x) - f(x)$. Thus at step $t\ge 1$ of Gradient Boosting, a CART regression tree is used to estimate the residual function $f_{\bm X, t-1}-f$. In practice other loss functions $\psi$ could be considered. The only requirement is that $\psi$ be differentiable with respect to its second variable, see \cite{fri_2001}.\footnote{There are several popular open-source software libraries that provide implementations of Gradient Boosting \cite{fri_2001}. The most popular ones are XGBoost (for eXtreme Gradient Boosting) \cite{che_gue_2016}, LightGBM (for Light Gradient Boosting Machine) \cite{guo_et_al_2017}, and CatBoost (for Categorical Boosting) \cite{pro_et_al_2018}. Since these libraries are based on several engineering optimizations, they provide estimators that are not exactly as $f_{\bm X, t}$ in \eqref{gradient_descent}. In these libraries, there is a wide range of loss functions available, and it is also possible to implement one's own loss function.}

%Algorithm \ref{grad_boost_algo} gives a pseudo-code to construct Gradient Boosting, where the number of boosting iterations is a hyperparameter.

% As discussed in the introductory part of this section, instead of CART other decision tree algorithms could be used for estimating the gradient in \eqref{gradient_descent}, see, e.g., C4.5 in Quinlan \cite{qui_1993} and Extremely randomized tree in Geurts et al. \cite{geu_et_al_2006}.

When to stop increasing the number of boosing iterations $t$? A standard approach to find the optimal $t$ is to use early stopping techniques, which is what we do in Section \ref{secexamples}. However, in practice Gradient Boosting is known to be resistant to overfitting, see, e.g., \cite{sch_etal_1997}. This means that, in general, the $L^2_\Q$-error $\|f_{\bm X, t}-f\|_{2,\Q}$, does not increase as $t$ becomes very large. 
%a large number of boosting iterations $t$ does not affect much the resulting estimator performance measured by the $L^2_\Q$-error $\|f_{\bm X}-f\|_{2,\Q}$, and estimated by Monte Carlo simulation using a large independent test sample, see Section \ref{secexamples} below.

Henceforth and throughout, $f_{\bm X}$ is a placeholder for either the Random Forest $f_{\bm X, \bm \Pi}$ in \eqref{rand_forest}, or the Gradient Boosting $f_{\bm X, t}$ in \eqref{gradient_descent}. The generic expression of $f_{\bm X}$ is given in \eqref{generic_el_estimator}.

% For the sake of presentation we denote the squared-error loss function by $\psi$, i.e., $\psi(x,y) = (x-y)^2$, so that $\E_{\Q}[(f(X)-h(X))^2] = \E_{\Q}[\psi(f(X), h(X))]$, for every $h \in L^2_\Q$.

% To introduce Gradient Boosting we start with the fact that $f(x)$ can be written as $f(x) = \arg\min_{c\in \R} \psi(f(x), c)$, $x\in E$. Therefore we can approximate $f$ using a gradient descent technique. Namely, we define the sequence of functions $(f_t)_{t\ge 0}$ recursively,
% \begin{equation}\label{theo_seq}
% \textstyle
% f_{t}(x) = f_{t-1}(x) - \gamma \partial_y\psi(f(x), f_{t-1}(x)), \quad x \in E,    
% \end{equation}
% where $f_0$ is an initial guess and $\gamma$ is a constant parameter called the step size.
% Using that $\psi(x, y) = (x-y)^2$, straightforward rewriting gives that $f_{t}(x) = (1-2\gamma)^{t} (f_0(x) - f(x)) + f(x)$. This shows that whenever $0<\gamma<1/2$ the function $f_t$ converges to $f$.

% In practice the function $f$ is only known at the training sample $\bm X$ and thus $f_t(x)$ in \eqref{theo_seq} is defined only for $x \in \bm X$. 

\section{Closed-form estimators for $V$}\label{section_closed_form_Veu_Vus}
In the previous section we presented the construction of an ensemble estimator $f_{\bm X}$ of $f$ of the form \eqref{generic_el_estimator}. Now we use $f_{\bm X}$ to define an estimator $V_{\bm X}$ of $V$ of the form \eqref{eqYhat}. From \eqref{generic_el_estimator} and \eqref{eqYhat}, we derive the following expression
\begin{equation}\label{hat_Vt_closed_form}
\textstyle
    V_{\bm X, t} = \sum_{i=1}^N \beta_i \E_\Q[ \1_{\bm A_i}(X) \mid \Fcal_t], \quad t = 0, \dots, T.
\end{equation}
Now recall that $X$ has distribution $\Q(dx) = \Q_1(dx_1) \times \dots \times\Q_T(dx_T)$. Thus for a hyperrectangle $\bm A=\prod_{t=1}
^T (a_t, b_t]$ in \eqref{box_notation}, we have\footnote{For $t=0$, we set $\prod_{s=1}^0\cdot=1$.}
\begin{equation}\label{prob_decomp}
\textstyle \E_\Q[ \1_{\bm A}(X) \mid \Fcal_t] = \prod_{s=1}^t\1_{( a_{s},  b_{s}]}(X_s)\prod_{s=t+1}^T\Q_s[( a_{s}, b_{s}]].
\end{equation}
Accordingly, we deduce that the value process estimator $V_{\bm X}$ is in closed form as soon as the probability
\begin{equation}\label{closed_form_Qs}
\Q_s[( a_{s},b_{s}]] \text{ is in closed form for every } a_s < b_s \in \overline{\R^d}, \, s=1, \dots, T.
\end{equation}
We say an expression is in closed form if it can be efficiently evaluated at very low computational cost. Below, we present two common cases where property \eqref{closed_form_Qs} is satisfied.

\subsection{Cross-sectional independence}\label{subsec_Qs_as_prod}
The first case is when $\Q_s$ can be factorized as $\Q_s(dx_s) = \Q_{1, s}(dx_{1,s}) \times \dots \times \Q_{d,s}(dx_{d,s})$. In this case, for $ a_s < b_s \in \overline{\R^d}$, we have $\Q_s[( a_{s},  b_{s}]] = \Q_s[ \prod_{j=1}^d ( a_{j, s}, b_{j, s}]] = \prod_{j=1}^d (F_{j, s}(b_{j, s}) - F_{j, s}(a_{j, s}))$, where $F_{j, s}$ denotes the cumulative distribution function of $\Q_{j, s}$. Thus property \eqref{closed_form_Qs} holds as soon as $F_{j,s}$ is in closed form for every $j=1, \dots, d$. There are many such examples. For its extensive use in financial modelling we mention the standard normal distribution $\Q_s=\Ncal(0, I_d)$. Examples include the discrete-time multivariate Black--Scholes model in \eqref{bs_model} and many more time-series models, such as the GARCH models in \cite{boll_1986}.

\subsection{Closed-form copulas}\label{subsec_Qs_with_copula}

The second case, generalizing the above, is when $\Q_{s}$ is defined in terms of a copula. As above, we denote by $F_{j, s}$ the cumulative distribution function of the marginal $\Q_{j, s}$. We now assume as given a copula $C_s$ on $[0,1]^d$ such that
\[ \textstyle    \Q_s[(-\infty,x_s]]= C_s(F_{1,s}(x_{1,s}), \dots, F_{d,s}(x_{d,s})).\]
In fact, it is well known that any multivariate distribution on $\R^d$ can be expressed in terms of a copula, see \cite{skl_1959} and \cite[Theorem~1]{emb_2009}. Moreover, the copula is unique if the marginals $F_{j, s}$ are continuous.

Now property \eqref{closed_form_Qs} holds as soon as the copula $C_s$ and the marginals $F_{j,s}$ are in closed form. Indeed, for any hyperrectangle $(a,b]=\prod_{j=1}^d(a_j,b_j]$ of $\R^d$, we have
\[
\textstyle\Q_s[(a,b]] = \sum_{z\in \prod_{j=1}^d \{  a_{j},  b_{j}\}} (-1)^{N(z)} C_s(F_{1,s}(z_1), \dots, F_{d,s}(z_d)),\quad N(z) =\mathrm{card}(\{k\mid z_k =   a_{k}\}).
\]
The first case corresponds to the independence copula $C_s(y)=\prod_{j=1}^d y_j$. Copula models are widespread in financial risk management, as they allow to design tailor-made dependence structures between the underlying assets. See, e.g., \cite{mcn_et_al_2015} for a thorough discussion.

\section{Numerical experiments}\label{secexamples}

We follow up on the introductory example with the Black–Scholes model with $d$ nominal stock price processes $S_{i, t}$ given by \eqref{bs_model}. In particular, we assume that $X_t$ are i.i.d.\ standard normal on $\R^d$.

As for the portfolios, we fix a strike price $K$ and consider the following European style exotic options with payoff functions
\begin{itemize}
\item Min-put $f(X)=  \e^{-r\sum_{t=1}^T \Delta_t} (K-\min_{i} S_{i,T})^+$;
\item Max-call $f(X)=\e^{-r\sum_{t=1}^T \Delta_t} (\max_{i} S_{i,T} -  K)^+$.
\end{itemize}
We also consider a genuinely path-dependent product with the payoff function
\begin{itemize}
\item Barrier reverse convertible (BRC) $f(X)= \e^{-r\sum_{t=1}^T \Delta_t}\left( C  +  F  \left( 1 - 1_{\{\min_{i,t}   S_{i,t}\le B\}}   \left(1 - \min_{i} \frac{ S_{i,T}}{S_{i,0}  K} \right)^+\right)\right)$,
\end{itemize}
for some barrier $B<K$, coupon $C$, and face value $F$. At maturity $T$, the holder of this structured product receives the coupon $C$. She also receives the face value $F$ if none of the nominal stock prices falls below the barrier $B$ at any time $t=1,\dots,T$. Otherwise, the face value $F$ is reduced by the payoff of $F/K$ min-puts on the normalized stocks $S_{i,T}/S_{i,0}$ with strike price $K$. These payoff functions are inspired from those given in \cite{bec_et_al_2019}. Note that the payoff functions of the min-put and BRC are bounded, while the payoff of the max-call is unbounded.

For our numerical experiments we choose the following parameter values: risk-free rate $r=0$, initial stock prices $S_{i, 0}=1$, volatilities $\sigma_i = 0.2\bm e_i$, where $\bm e_i$ denote the standard basis vectors in $\R^d$, so that stock prices are independent, strike price $K=1$ (at the money), barrier $B=0.6$, coupon $C=0$, and face value $F=1$. For the min-put and max-call, $(d, T) = (6, 2)$ and $(\Delta_1, \Delta_2) = (1/12, 11/12)$; for the BRC, $(d, T) = (3, 12)$ and $(\Delta_1, \dots, \Delta_{12}) = (1/12, \dots, 1/12)$. Thus the path space $\R^{d\times T}$ is of dimension $12$ for the min-put and max-call, and it is of dimension $36$ for the BRC.

Under the parameter specification above, we generate a training sample $\bm X$ of size $n= 20{,}000$. We use $\bm X$, along with the corresponding function values $\bm f$, to construct the ensemble estimator $f_{\bm X}$ in \eqref{generic_el_estimator}. To find the optimal hyperparameter value for this estimator we use a validation sample $\bm X_{\mathrm{valid}}$ of size $0.4\times n=8{,}000$, along with its corresponding function values $\bm f_{\mathrm{valid}}$. Both the optimal hyperparameter value search and the construction of $f_{\bm X}$ are done using the programming language Python and readily accessible machine learning libraries. 

Specifically, when $f_{\bm X}$ is the Random Forest $f_{\bm X, \bm\Pi}$ in \eqref{rand_forest}, we use the RandomForestRegressor class of the library scikit-learn \cite{scikit_learn}. We find the optimal hyperparameter value by validation on the set of hyperparameter values $\Pcal_{\mathrm{RF}} =\{(M, \textbf{nodesize}, p)\mid M \in \{100, 250, 500\},\, \textbf{nodesize}\in\{2, 3, 5\},\, p \in \{\lceil d\times T/3\rceil, d\times T\} \}$ using $\bm X_{\mathrm{valid}}$ and $\bm f_{\mathrm{valid}}$. In $\Pcal_{\mathrm{RF}}$ there are three default hyperparameter values. The RandomForestRegressor (Python) default hyperparameter value $(100, 2, d\times T)$, the randomForest \cite{rf_r_pack} (R programming language) default hyperparameter value in regression $(500, 5, \lceil d\times T/3\rceil)$, and our default hyperparameter value $(100, 5, d\times T)$.\footnote{Our default hyperparameter value is an intermediary choice between the default hyperparameter values in RandomForestRegressor and randomForest.} Table \ref{table_parameters_rf} shows the normalized $L^2_\Q$-error $\|f_{\bm X}-f\|_{2,\Q}/V_0$, computed using the validation sample $\bm X_{\mathrm{valid}}$, and the number of hyperrectangles $N$ in the Random Forest $f_{\bm X}$ in \eqref{generic_el_estimator} for these three default hyperparameter values as well as the optimal hyperparameter value in $\Pcal_{\mathrm{RF}}$. For the min-put and max-call, we observe that our default hyperparameter value gives normalized $L^2_\Q$-error comparable to that given by the optimal hyperparameter value. Besides it has the advantage to give, on average, 8 times less hyperrectangles than the optimal hyperparameter value. This implies that the evaluation of $V_{\bm X}$ is 8 times faster with our default hyperparameter value than with the optimal hyperparameter value for the min-put and max-call examples. Thus for computational reason we use our default hyperparameter value $(100,5, 12)$ for min-put and max-call. However for BRC we use the optimal hyperparameter value $(500, 5, 12)$, because here the number of hyperrectangles is relatively small ($N<500{,}000$). For the three payoff functions we use $\textbf{sampling regime}=$bootstrapping. In the class RandomForestRegressor, the variables $M$, \textbf{nodesize}, $p$, \textbf{sampling regime} correspond to \textbf{n\_estimators}, \textbf{min\_samples\_split}, \textbf{max\_features}, \textbf{bootstrap}, respectively.

\begin{table}
\centering
  \begin{tabular}{|l|l|l|l|}
\hline
 & Min-put  & BRC & Max-call     \\
\hline
Optimal hyperparameter value & (500, 2, 12)  & (500, 5, 12) &(250, 3, 12)   \\
Normalized $L^2_\Q$-error in \% &6.864 & 6.884 & 10.26  \\
Number of hyperrectangles &6{,}279{,}290 & 187{,}710 & 2{,}027{,}347 \\
\hline
Default hyperparameter value in RandomForestRegressor (Python) & (100, 2, 12)  & (100, 2, 36) &(100, 2, 12) \\
Normalized $L^2_\Q$-error in \% & 6.894 & 6.973 & 10.36 \\
Number of hyperrectangles &1{,}255{,}344 & 52{,}805 & 1{,}237{,}737 \\
\hline
Default hyperparameter value in randomForest (R) & (500, 5, 4)  & (500, 5, 12) &(500, 5, 4) \\
Normalized $L^2_\Q$-error in \% & 8.124 & 6.884 & 12.61 \\
Number of hyperrectangles &2{,}575{,}215 & 187{,}710 & 2{,}556{,}448 \\
\hline
\textbf{Our default hyperparameter value} &(100, 5, 12)  & (100, 5, 36) &(100, 5, 12) \\
Normalized $L^2_\Q$-error in \% & 6.917& 6.965  & 10.39 \\
Number of hyperrectangles &494{,}118 & 34{,}807 & 489{,}747 \\
\hline
\end{tabular}
\caption{Random Forest validation step: normalized $L^2_\Q$-error $\|f_{\bm X}-f\|_{2, \Q}/V_0$, computed using the validation sample $\bm X_{\mathrm{valid}}$ and expressed in \%, and number of hyperrectangles $N$ in the Random Forest $f_{\bm X}$ in \eqref{generic_el_estimator}, for the optimal hyperparameter value in $\Pcal_{\mathrm{RF}}$, and three default hyperparameter values, for the payoff functions min-put, BRC, and max-call.}\label{table_parameters_rf}
\end{table}

When $f_{\bm X}$ is the Gradient Boosting $f_{\bm X, t}$ in \eqref{gradient_descent}, we use the XGBRegressor class of XGBoost \cite{che_gue_2016}. Similarly to what we did for Random Forest, we use $\bm X_{\mathrm{valid}}$ and $\bm f_{\mathrm{valid}}$ to perform a validation on the set of hyperparameter values $\Pcal_{\mathrm{XGB}}=\{(t_{\mathrm{optimal\_stopping}}(\textbf{nodesize}, \textbf{max depth}),\textbf{nodesize}, \textbf{max depth})\mid \textbf{nodesize} \in \{5, 15, 25, 35, 45\},\, \textbf{max depth} \in  \{40, 50, \dots, 90\} \}$ to find the optimal hyperparameter value. The hyperparameter \textbf{max depth} controls the number of hyperrectangles in the regression tree $g_t$ in \eqref{gradient_descent}. Given the values \textbf{nodesize}, \textbf{max depth}, the number of iterations $t_{\mathrm{optimal\_stopping}}(\textbf{nodesize}, \textbf{max depth})$ is determined by early stopping using the validation sample $\bm X_{\mathrm{valid}}$. Table \ref{table_parameters_xgb} shows the normalized $L^2_\Q$-error $\|f_{\bm X}-f\|_{2,\Q}/V_0$, computed using the validation sample $\bm X_{\mathrm{valid}}$, and the number of hyperrectangles $N$ in the Gradient Boosting $f_{\bm X}$ in \eqref{generic_el_estimator} for the optimal hyperparameter value in $\Pcal_{\mathrm{XGB}}$. Furthermore, for these three payoff functions we also considered other hyperparameters in XGBRegressor for which we took standard values: $\textbf{booster}=\text{gbtree}$, $\textbf{learning\_rate}=0.1$, $\textbf{tree\_method}=\text{hist}$, $\textbf{objective}=\text{reg:squarederror}$, and $\textbf{base\_score}=0.5$. Note that in XGBRegressor, the variables \textbf{nodesize} and \textbf{max depth} correspond to \textbf{min\_child\_weight} and \textbf{max\_depth}, respectively.

\begin{table}
\centering
  \begin{tabular}{|l|l|l|l|}
\hline
 & Min-put  & BRC & Max-call     \\
\hline
Optimal hyperparameter value & (120, 40, 15)  & (256, 50, 15) &(152, 60, 35)   \\
Normalized $L^2_\Q$-error in \% &5.856 & 6.284 & 9.753 \\
Number of hyperrectangles &88{,}127 & 189{,}864 & 66{,}225 \\
\hline
\end{tabular}
\caption{XGBoost validation step: normalized $L^2_\Q$-error $\|f_{\bm X}-f\|_{2, \Q}/V_0$, computed using the validation sample $\bm X_{\mathrm{valid}}$ and expressed in \%, and number of hyperrectangles $N$ in the Gradient Boosting $f_{\bm X}$ in \eqref{generic_el_estimator}, for the optimal hyperparameter value in $\Pcal_{\mathrm{XGB}}$, for the payoff functions min-put, BRC, and max-call.}\label{table_parameters_xgb}
\end{table}

Next we use our ensemble estimator $f_{\bm X}$ to construct $V_{\bm X}$ in \eqref{eqYhat}. As discussed in Section \ref{section_closed_form_Veu_Vus}, $V_{\bm X}$ is given in closed form. We then evaluate $V_{\bm X,t}$ at times $t \in \{0, 1, T\}$ on a test sample $\bm X_{\mathrm{test}}$ of size $n_{\mathrm{test}} = 100{,}000$. We benchmark $V_{\bm X}$ to the ground truth value process $V$, which we obtain by means of Monte Carlo schemes using $\bm X_{\mathrm{test}}$. More specifically, we obtain $V_0$ as simple Monte Carlo estimate of $\{f(X)\mid X \in \bm X_{\mathrm{test}}\}$. For $V_1$, we use a nested Monte Carlo scheme, where we estimate $V_1(X_1)$ using $n_{\mathrm{inner}} = 1{,}000$ inner simulations of $(X_2,\dots,X_T)$, for each $X_1$ in $\bm X_{\mathrm{test}}$. Then we carry out the following three evaluation tasks.

\begin{figure}[p]
    \centering % <-- added
    \begin{subfigure}{0.45\textwidth}
  \includegraphics[width=\linewidth]{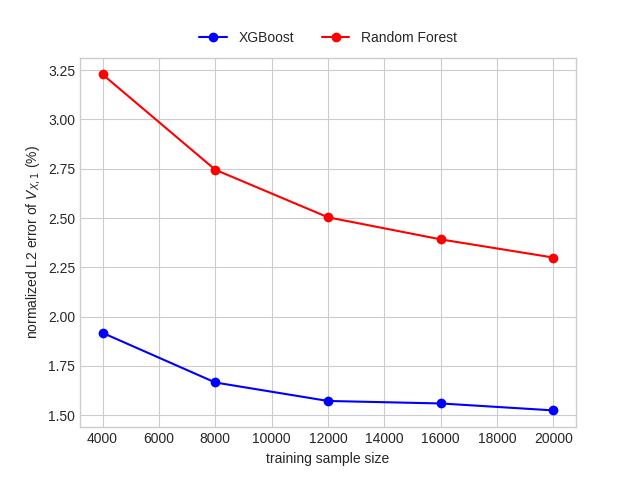}
  \caption{Normalized $L^2_{\Q}$-errors of $V_{\bm X, 1}$ in \% with Random Forest and XGBoost.}
  \label{error_V1_minput_rho0}
\end{subfigure}\hfil % <-- added
\begin{subfigure}{0.45\textwidth}
  \includegraphics[width=\linewidth]{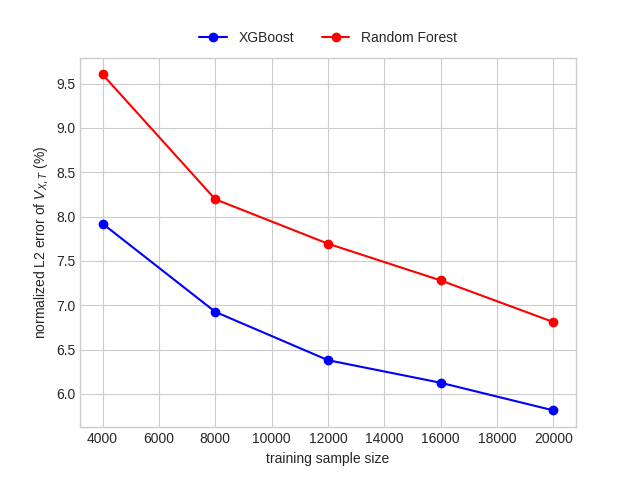}
  \caption{Normalized $L^2_{\Q}$-errors of $V_{\bm X, T}$ in \% with Random Forest and XGBoost.}
  \label{error_VT_minput_rho0}
\end{subfigure}
\medskip
\begin{subfigure}{0.45\textwidth}
  \includegraphics[width=\linewidth]{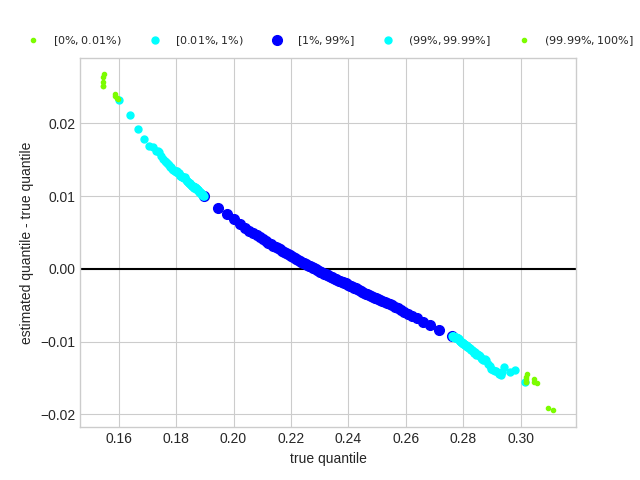}
  \caption{Detrended Q-Q plot of $V_{\bm X, 1}$ with Random Forest.}
  \label{qqplot_V1_minput_rf}
\end{subfigure}\hfil % <-- added
\begin{subfigure}{0.45\textwidth}
  \includegraphics[width=\linewidth]{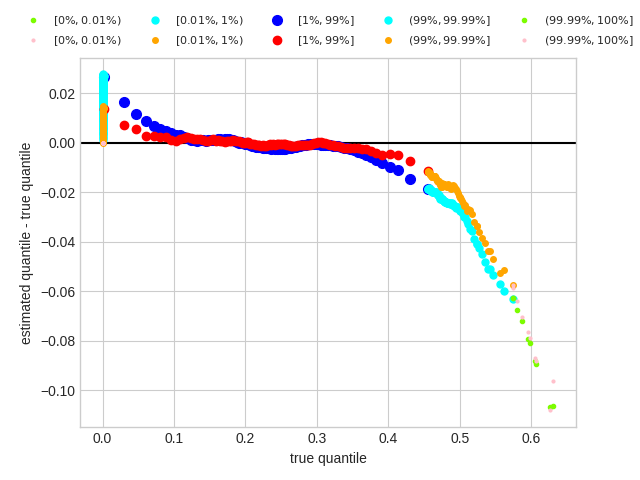}
  \caption{Detrended Q-Q plot of $V_{\bm X, T}$ with Random Forest.}
  \label{qqplot_VT_minput_rf}
\end{subfigure}
\medskip
\begin{subfigure}{0.45\textwidth}
  \includegraphics[width=\linewidth]{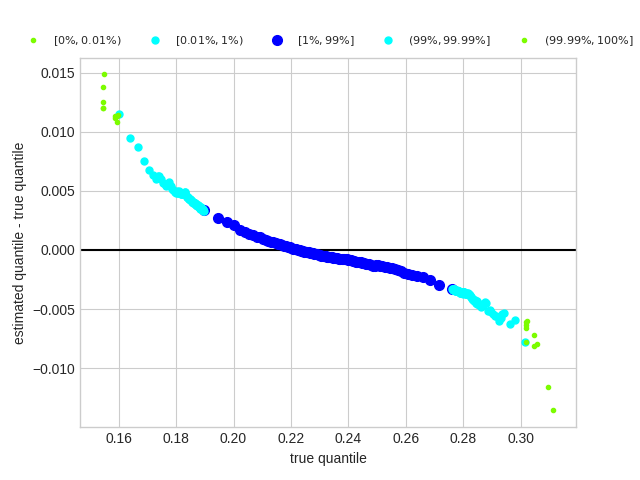}
  \caption{Detrended Q-Q plot of $V_{\bm X, 1}$ with XGBoost.}
  \label{qqplot_V1_minput_xgb}
\end{subfigure}\hfil % <-- added
\begin{subfigure}{0.45\textwidth}
  \includegraphics[width=\linewidth]{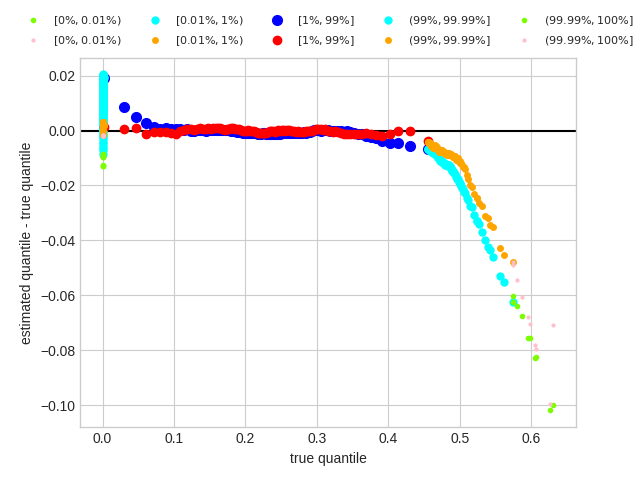}
  \caption{Detrended Q-Q plot of $V_{\bm X, T}$ with XGBoost.}
  \label{qqplot_VT_minput_xgb}
\end{subfigure}
\caption{Results for the min-put with Random Forest and XGBoost. The normalized $L^2_\Q$-error of $V_{\bm X, 1}$, $\|V_1-V_{\bm X, 1}\|_{2, \Q}/V_0$, is computed using the test sample and expressed in \%. In the detrended Q-Q plots, the blue, cyan, and lawngreen (red, orange, and pink) dots are built using the test (training) sample. $[0\%, 0.01\%)$ refers to the quantiles of levels $\{0.001\%,0.002\%, \dots, 0.009\%\}$, $[0.01\%, 1\%)$ refers to the quantiles of levels $\{0.01\%,0.02\%, \dots, 0.99\%\}$, $[1\%, 99\%]$ refers to the quantiles of levels $\{1\%,2\%, \dots, 99\%\}$, $(99\%, 99.99\%]$ refers to the quantiles of levels $\{99.01\%,99.02\%, \dots, 99.99\%\}$, and $(99.99\%, 100\%]$ refers to the quantiles of levels $\{99.991\%,99.992\%, \dots, 100\%\}$.}
\label{fig_minput}
\end{figure}

\begin{figure}[p]
    \centering % <-- added
    \begin{subfigure}{0.45\textwidth}
  \includegraphics[width=\linewidth]{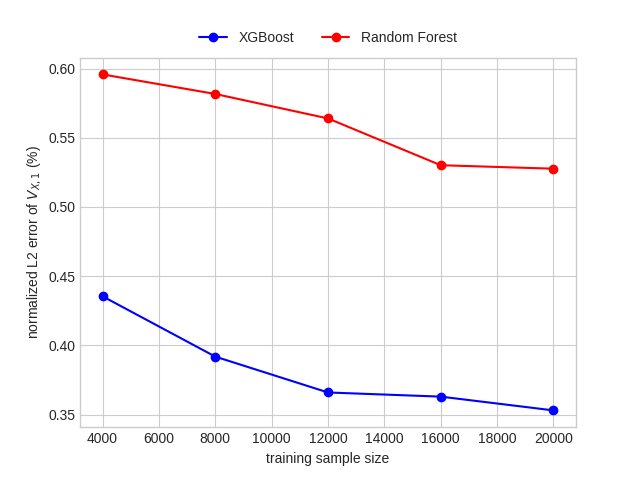}
  \caption{Normalized $L^2_{\Q}$-errors of $V_{\bm X, 1}$ in \% with Random Forest and XGBoost.}
  \label{error_V1_barrier_rho0}
\end{subfigure}\hfil % <-- added
\begin{subfigure}{0.45\textwidth}
  \includegraphics[width=\linewidth]{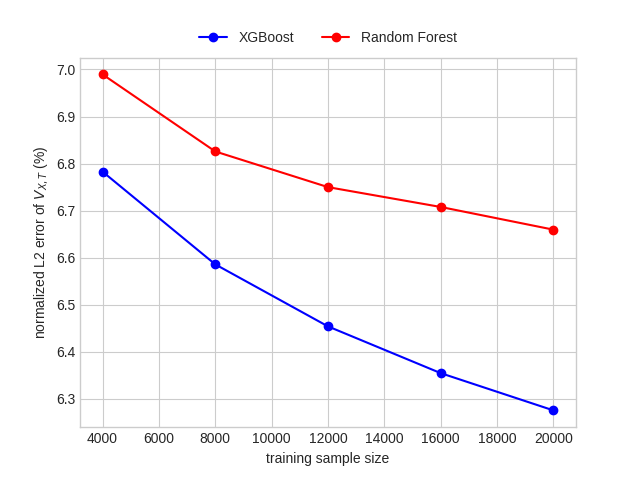}
  \caption{Normalized $L^2_{\Q}$-errors of $V_{\bm X, T}$ in \% with Random Forest and XGBoost.}
  \label{error_VT_barrier_rho0}
\end{subfigure}
\medskip
\begin{subfigure}{0.45\textwidth}
  \includegraphics[width=\linewidth]{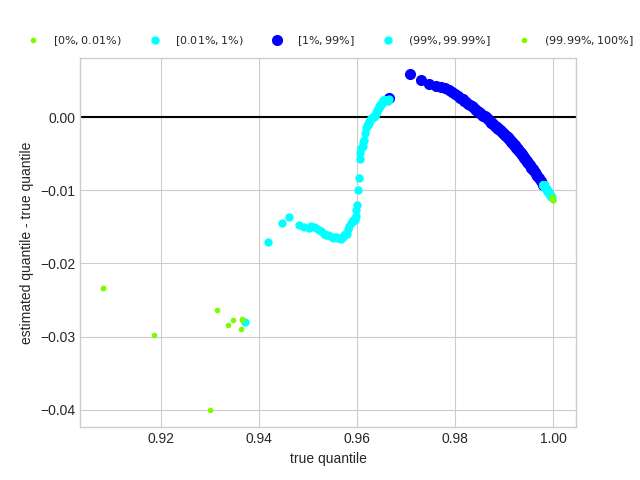}
  \caption{Detrended Q-Q plot of $V_{\bm X, 1}$ with Random Forest.}
  \label{qqplot_V1_barrier_rf}
\end{subfigure}\hfil % <-- added
\begin{subfigure}{0.45\textwidth}
  \includegraphics[width=\linewidth]{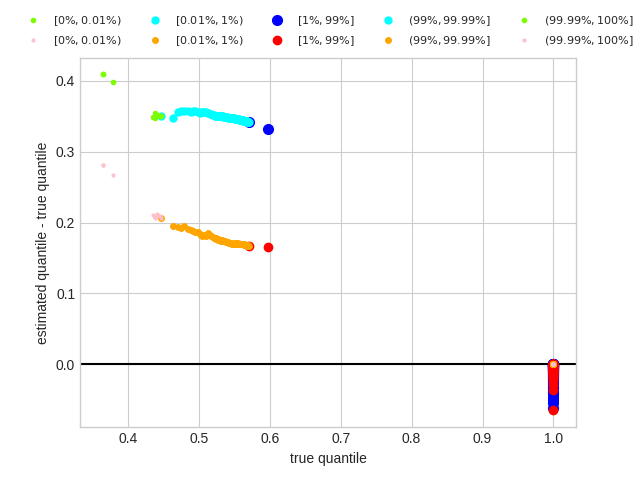}
  \caption{Detrended Q-Q plot of $V_{\bm X, T}$ with Random Forest.}
  \label{qqplot_VT_barrier_rf}
\end{subfigure}
\medskip
\begin{subfigure}{0.45\textwidth}
  \includegraphics[width=\linewidth]{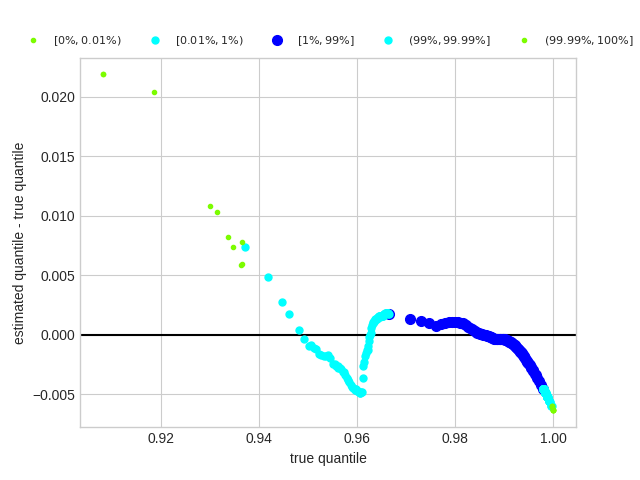}
  \caption{Detrended Q-Q plot of $V_{\bm X, 1}$ with XGBoost.}
  \label{qqplot_V1_barrier_xgb}
\end{subfigure}\hfil % <-- added
\begin{subfigure}{0.45\textwidth}
  \includegraphics[width=\linewidth]{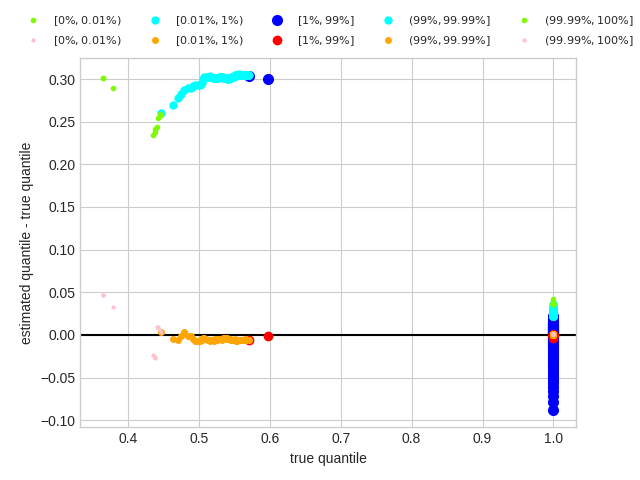}
  \caption{Detrended Q-Q plot of $V_{\bm X, T}$ with XGBoost.}
  \label{qqplot_VT_barrier_xgb}
\end{subfigure}
\caption{Results for the BRC with Random Forest and XGBoost. The normalized $L^2_\Q$-error of $V_{\bm X, 1}$, $\|V_1-V_{\bm X, 1}\|_{2, \Q}/V_0$, is computed using the test sample and expressed in \%. In the detrended Q-Q plots, the blue, cyan, and lawngreen (red, orange, and pink) dots are built using the test (training) sample. $[0\%, 0.01\%)$ refers to the quantiles of levels $\{ 0.001\%,0.002\%, \dots, 0.009\%\}$, $[0.01\%, 1\%)$ refers to the quantiles of levels $\{0.01\%,0.02\%, \dots, 0.99\%\}$, $[1\%, 99\%]$ refers to the quantiles of levels $\{1\%,2\%, \dots, 99\%\}$, $(99\%, 99.99\%]$ refers to the quantiles of levels $\{99.01\%,99.02\%, \dots, 99.99\%\}$, and $(99.99\%, 100\%]$ refers to the quantiles of levels $\{99.991\%,99.992\%, \dots, 100\%\}$.}
\label{fig_barrier}
\end{figure}

\begin{figure}[p]
    \centering % <-- added
    \begin{subfigure}{0.45\textwidth}
  \includegraphics[width=\linewidth]{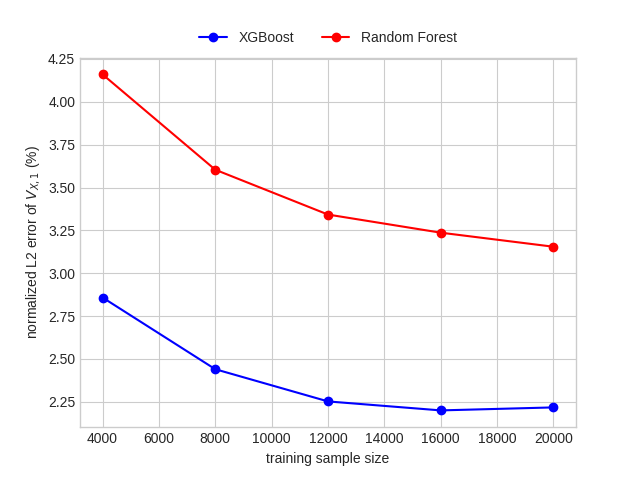}
  \caption{Normalized $L^2_{\Q}$-errors of $V_{\bm X, 1}$ in \% with Random Forest and XGBoost.}
  \label{error_V1_maxcall_rho0}
\end{subfigure}\hfil % <-- added
\begin{subfigure}{0.45\textwidth}
  \includegraphics[width=\linewidth]{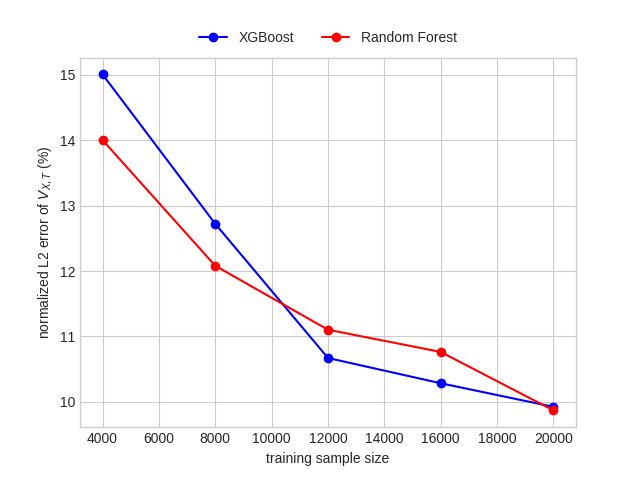}
  \caption{Normalized $L^2_{\Q}$-errors of $V_{\bm X, T}$ in \% with Random Forest and XGBoost.}
  \label{error_VT_maxcall_rho0}
\end{subfigure}
\medskip
\begin{subfigure}{0.45\textwidth}
  \includegraphics[width=\linewidth]{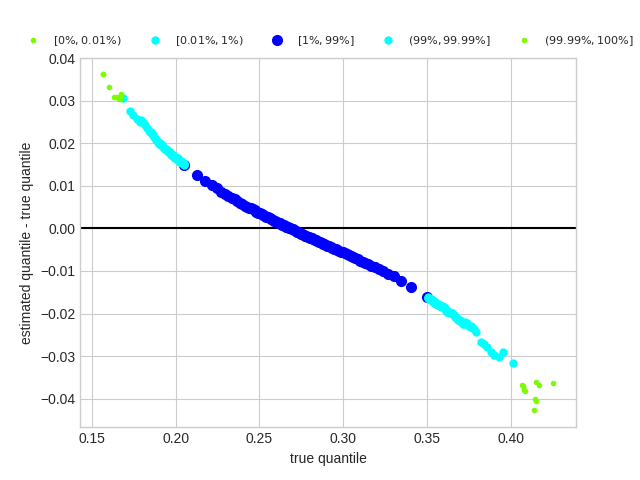}
  \caption{Detrended Q-Q plot of $V_{\bm X, 1}$ with Random Forest.}
  \label{qqplot_V1_maxcall_rf}
\end{subfigure}\hfil % <-- added
\begin{subfigure}{0.45\textwidth}
  \includegraphics[width=\linewidth]{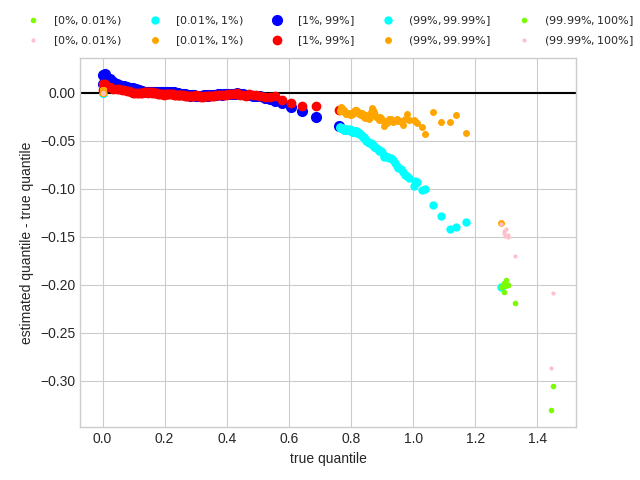}
  \caption{Detrended Q-Q plot of $V_{\bm X, T}$ with Random Forest.}
  \label{qqplot_VT_maxcall_rf}
\end{subfigure}
\medskip
\begin{subfigure}{0.45\textwidth}
  \includegraphics[width=\linewidth]{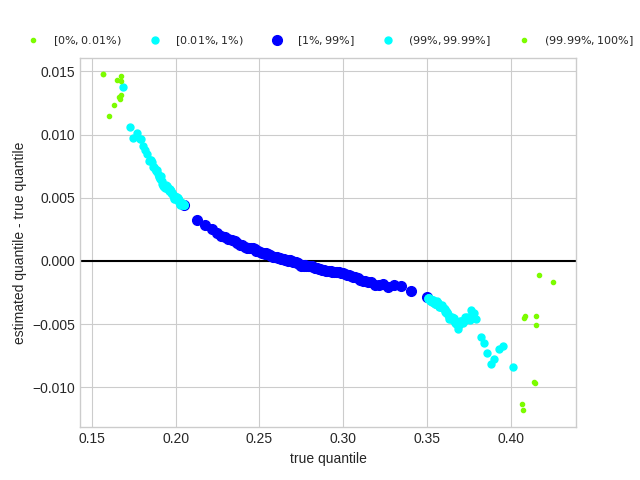}
  \caption{Detrended Q-Q plot of $V_{\bm X, 1}$ with XGBoost.}
  \label{qqplot_V1_maxcall_xgb}
\end{subfigure}\hfil % <-- added
\begin{subfigure}{0.45\textwidth}
  \includegraphics[width=\linewidth]{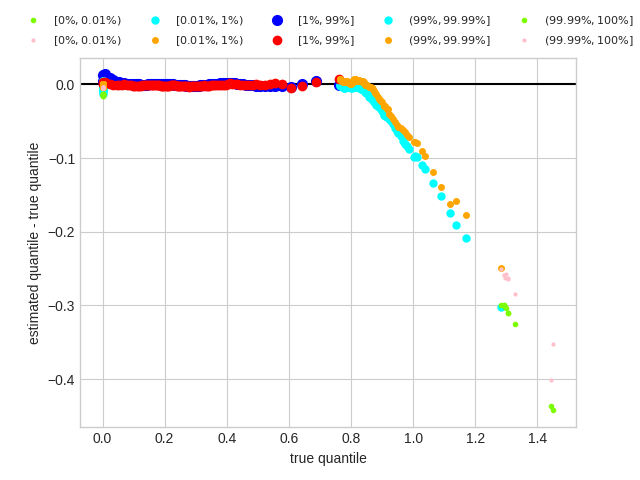}
  \caption{Detrended Q-Q plot of $V_{\bm X, T}$ with XGBoost.}
  \label{qqplot_VT_maxcall_xgb}
\end{subfigure}
\caption{Results for the max-call with Random Forest and XGBoost. The normalized $L^2_\Q$-error of $V_{\bm X, 1}$, $\|V_1-V_{\bm X, 1}\|_{2, \Q}/V_0$, is computed using the test sample and expressed in \%. In the detrended Q-Q plots, the blue, cyan, and lawngreen (red, orange, and pink) dots are built using the test (training) sample. $[0\%, 0.01\%)$ refers to the quantiles of levels $\{0.001\%,0.002\%, \dots, 0.009\%\}$, $[0.01\%, 1\%)$ refers to the quantiles of levels $\{0.01\%,0.02\%, \dots, 0.99\%\}$, $[1\%, 99\%]$ refers to the quantiles of levels $\{1\%,2\%, \dots, 99\%\}$, $(99\%, 99.99\%]$ refers to the quantiles of levels $\{99.01\%,99.02\%, \dots, 99.99\%\}$, and $(99.99\%, 100\%]$ refers to the quantiles of levels $\{99.991\%,99.992\%, \dots, 100\%\}$.}
\label{fig_maxcall}
\end{figure}

First, we compute the absolute relative error of $V_{\bm X, 0}$, $|V_0 - V_{\bm X, 0}|/V_0$, and the normalized $L^2_{\Q}$-errors of $V_{\bm X, t}$, $\|V_t - V_{\bm X, t}\|_{2, \Q}/V_0$, for $t=1, T$. Table \ref{norm_l2_errors_table} shows that normalized $L^2_\Q$-error of $V_{\bm X, t}$ decreases substantially for increasing time-to-maturity $T-t$. More specifically, for XGBoost, the normalized $L^2_\Q$-error of $V_{\bm X, 1}$ is on average 9-times smaller than that of $V_{\bm X, T}$, and the relative absolute error of $V_{\bm X, 0}$ is on average 14-times smaller than the normalized $L^2_\Q$-error of $V_{\bm X, 1}$. For Random Forest these values are 6 and 6, respectively. These findings are in line with \eqref{doobineq}, which has useful practical implications. Indeed, despite the lack of theoretical bounds on the error $\|V_t-V_{\bm X, t}\|_{2, \Q}$, in concrete applications one can always estimate the normalized $L^2_{\Q}$-error of $V_{\bm X, T}$ by a simple Monte Carlo scheme as we do here. This error then serves as upper bound on the normalized $L^2_{\Q}$-errors of $V_{\bm X, t}$, for any $t<T$. Table \ref{norm_l2_errors_table} also reveals that XGBoost outperforms Random Forest in the estimation of $V_{0}$, $V_{1}$ and $V_{T}$ in most cases (8 cases out of 9). In this table we also report the normalized $L^2_\Q$-errors obtained with the kernel-based method in our previous paper \cite[Table~2]{bou_fil_22}. We see that the kernel-based method outperforms our ensemble learning method only in the BRC example. Figures \ref{error_V1_minput_rho0}, \ref{error_V1_barrier_rho0}, and \ref{error_V1_maxcall_rho0} show the decrease of the normalized $L^2_\Q$-error of $V_{\bm X, 1}$ with respect to the training sample size $n$. Figures \ref{error_VT_minput_rho0}, \ref{error_VT_barrier_rho0}, and \ref{error_VT_maxcall_rho0} illustrate the same phenomenon for $V_{\bm X, T}$. In these figures we also recognize the outperformance of XGBoost over Random Forest.

Second, we compute and compare quantiles of $V_{\bm X, 1}$ and $V_1$, and $V_{\bm X, T}$ and $V_T$ using the test and training sample. Thereto, for $t\in \{1, T\}$, we compute the empirical left quantiles of $V_{\bm X, t}$ and $V_t$ at levels $\{0.001\%, 0.002\%, \dots, 0.009\%\}$, $\{0.01\%, 0.02\%,\dots, 0.99\%\}$, $\{1\%,2\%,\dots,99\%\}$, $\{99.01\%, 99.02\%, \dots, 99.99\%\}$, and $\{99.991\%, 99.992\%,\dots,  100\%\}$.\footnote{Note that for the test sample of size $n_{\mathrm{test}}=10^5$, the left $0.001\%$-quantile ($100\%$-quantile) corresponds to the smallest (largest) sample value. For the training sample of size $n_{\mathrm{train}}=2\times 10^4$, the same holds, while the ten left- and right-most quantiles collapse to two values, respectively.} The detrended quantiles (estimated quantiles minus true quantiles) are then plotted against the true quantiles, the produced plot is called a detrended Q-Q plot. Figures \ref{qqplot_V1_minput_rf}, \ref{qqplot_V1_barrier_rf}, \ref{qqplot_V1_maxcall_rf}, and Figures \ref{qqplot_V1_minput_xgb}, \ref{qqplot_V1_barrier_xgb}, \ref{qqplot_V1_maxcall_xgb} show the detrended Q-Q plots of $V_{\bm X, 1}$ with Random Forest and XGBoost, respectively. These figures show that overall the distribution of $V_{1}$ is better estimated with XGBoost than with Random Forest. Figures \ref{qqplot_VT_minput_rf}, \ref{qqplot_VT_barrier_rf}, \ref{qqplot_VT_maxcall_rf}, and Figures \ref{qqplot_VT_minput_xgb}, \ref{qqplot_VT_barrier_xgb}, \ref{qqplot_VT_maxcall_xgb} show the detrended Q-Q plots of $V_{\bm X, T}$ with Random Forest and XGBoost, respectively. Notably, the detrended Q-Q plots of $V_{\bm X, T}$ in Figures \ref{qqplot_VT_barrier_rf} and \ref{qqplot_VT_barrier_xgb} reveal that for less than $3\%$ of the training sample (that is, less than $600$ points out of $n= 20{,}000$), the embedded min-put options in the BRC are triggered and in the money. For the remaining sample points the payoff is equal to the face value, $F=1$. And yet, as Figures \ref{qqplot_V1_barrier_rf} and \ref{qqplot_V1_barrier_xgb} show, this is enough for our ensemble learning method to learn the payoff function such that $V_{\bm X, 1}$ is remarkably close to the ground truth, with a normalized $L^2_\Q$-error less than $0.60\%$, as reported in Table \ref{norm_l2_errors_table}. In \cite{bou_fil_22} we also compute the same detrended Q-Q plots as here. Overall, the detrended Q-Q plots drawn with the kernel-based method, see \cite[Figures~1-3]{bou_fil_22}, and those drawn with the ensemble learning method are of comparable quality.

Third, as risk management application, we compute the value at risk and expected shortfall of long and short positions of the above portfolios. Thereto, we recall the definitions that can also be found in \cite[Chapter~4]{foe_sch_04}. For a confidence level $\alpha \in (0,1)$ and random loss $\mathrm{L}$, the value at risk of $\mathrm{L}$ is defined as left $\alpha$-quantile $\mathrm{VaR}_\alpha(\mathrm{L})=\inf\{y \mid \Pa[\mathrm{L}\le y] \ge \alpha\}$, and the expected shortfall of $L$ is given by $\mathrm{ES}_{\alpha}(\mathrm{L}) = \frac{1}{1-\alpha}\E_\Pa[(\mathrm{L}-\mathrm{VaR}_\alpha(\mathrm{L}))^+] + \mathrm{VaR}_\alpha(\mathrm{L})$. Both value at risk and expected shortfall are standard risk measures in practice. For instance, insurance companies have to compute the value at risk at level $\alpha=99.5\%$ and the expected shortfall at level $\alpha=99\%$, under Solvency II and the Swiss Solvency Test, respectively. For more discussion on these two risk measures we refer to \cite{mcn_et_al_2015}. Henceforth, we assume the real-world measure $\Pa=\Q$, for simplicity. For the three payoff functions above, we compute value at risk and expected shortfall of the 1-period loss $\mathrm{L}=V_{0}-V_{1}$ and its estimator $\mathrm{L}_{\bm X}=V_{\bm X, 0}-V_{\bm X, 1}$ for a long position, namely $\mathrm{VaR}_{99.5\%}(\mathrm{L})$, $\mathrm{ES}_{99\%}(\mathrm{L})$, $\mathrm{VaR}_{99.5\%}(\mathrm{L}_{\bm X})$, and $\mathrm{ES}_{99\%}(\mathrm{L}_{\bm X})$. We compute the same risk measures for a short position, namely $\mathrm{VaR}_{99.5\%}(-\mathrm{L})$, $\mathrm{ES}_{99\%}(-\mathrm{L})$, $\mathrm{VaR}_{99.5\%}(-\mathrm{L}_{\bm X})$, and $\mathrm{ES}_{99\%}(-\mathrm{L}_{\bm X})$. And in Tables \ref{var_table} and \ref{es_table}, we report the relative errors of risk measures, namely $(\text{estimated risk measure minus true risk measure})/\text{true risk measure}$, which are expressed in \%. From these two tables, we notice that all risk measures are more accurately estimated with XGBoost than with Random Forest. This echoes the detrended Q-Q plots of $V_{\bm X, 1}$, where we see in Figures \ref{qqplot_V1_minput_rf}, \ref{qqplot_V1_barrier_rf}, \ref{qqplot_V1_maxcall_rf} and Figures \ref{qqplot_V1_minput_xgb}, \ref{qqplot_V1_barrier_xgb}, \ref{qqplot_V1_maxcall_xgb} that the left and right tails of the distribution of $V_1$ are better estimated with XGBoost than with Random Forest. With XGBoost the estimates of risk measures are satisfactory. In Tables \ref{var_table} and \ref{es_table}, we also compute the relative errors of risk measures with the kernel-based method using the risk measurements in \cite[Tables~3-4]{bou_fil_22}. We observe that risk measures are better estimated with the kernel-based method than with the ensemble learning method. Nevertheless, one should keep in mind that these risk measures are a tough metric for our estimators, because they focus on the tails of the distribution beyond the $1\%$- and $99\%$-quantiles. 

%This echoes the high quality of the detrended Q-Q plots in \cite[Figures~1-3]{bou_fil_22}, which stipulates that the kernel-based method is better at extrapolating the estimator $f_{\bm X}$ beyond the range of the sample $\bm X$ than the ensemble learning method.

% Other computations, not reported here, show the evolution of Root Mean Square Error (RMSE) during the training of regress-later XGBoost. They highlight the following well-known result in the machine learning literature: Gradient Boosting is very resisting to overfitting, in the sense that a large number of boosting iterations does not deteriorate the performance of the estimator.

%%% normalized errors table
\begin{table}
\centering
  \begin{tabular}{|l|l|l|l|l|}
\hline
Payoff & Estimator  &    $V_{\bm X, 0}$ &         $V_{\bm X, 1}$ &         $V_{\bm X, T}$ \\
\hline
Min-put & XGBoost & \textbf{0.1701} & \textbf{1.525} & \textbf{5.814}   \\
 & Random Forest & 0.2933 & 2.300 & 6.811   \\
  & Kernel-based method & 0.1942 & 1.827 & 10.05   \\
\hline
BRC & XGBoost & 0.05519 & 0.3530 & 6.276   \\
 & Random Forest & 0.2008 & 0.5276 & 6.660  \\
 & Kernel-based method & \textbf{0.02198}& \textbf{0.2506} & \textbf{5.745}   \\
\hline
Max-call & XGBoost & \textbf{0.08016} & \textbf{2.217} & 9.923 \\
 & Random Forest & 0.3845 & 3.155 & \textbf{9.868}  \\
 & Kernel-based method & 0.1031& 2.315& 11.65   \\
\hline
\end{tabular}
\caption{Normalized $L^2_\Q$-error $\|V_t-V_{\bm X, t}\|_{2, \Q}/V_0$, computed using the test sample and expressed in \%, at steps $t = 0,1,T$, using XGBoost and Random Forest, for the payoff functions min-put, BRC, and max-call.}\label{norm_l2_errors_table}
\end{table}

%%% VAR table
\begin{table}
\centering
  \begin{tabular}{|l|l|l|l|}
\hline
Payoff & Estimator &  $\mathrm{VaR}(\mathrm{L}_{\bm X})$& $\mathrm{VaR}(-\mathrm{L}_{\bm X})$\\
\hline
Min-put & XGBoost &  -9.658 & -6.912   \\
 & Random Forest &  -25.69 & -20.57   \\
 & Kernel-based method &  \textbf{0.9695}& \textbf{3.158}
   \\
\hline
BRC & XGBoost & 2.533 & -42.85   \\
 &  Random Forest &  -3.510 & -75.87   \\
 & Kernel-based method & \textbf{0.1893}& \textbf{-13.91}   \\
\hline
Max-call & XGBoost & -7.103 & -4.140   \\
 & Random Forest & -23.87 & -20.51   \\
 & Kernel-based method & \textbf{0.07143}& \textbf{-3.582}  \\
 \hline
\end{tabular}
\caption{
Relative errors of value at risk $\mathrm{VaR}_{99.5\%}(\mathrm{L}_{\bm X})$ and $\mathrm{VaR}_{99.5\%}(-\mathrm{L}_{\bm X})$, computed as $(\text{estimated VaR minus true VaR})/\text{true VaR}$ using the test sample and expressed in \%, using XGBoost and Random Forest, for the payoff functions min-put, BRC, and max-call.}\label{var_table}
\end{table}

%%% ES table
\begin{table}
\centering
  \begin{tabular}{|l|l|l|l|}
\hline
Payoff & Estimator &  $\mathrm{ES}(\mathrm{L}_{\bm X})$& $\mathrm{ES}(-\mathrm{L}_{\bm X})$\\
\hline
Min-put & XGBoost & -10.23 & -7.434   \\
 & Random Forest & -26.41 & -21.07   \\
 & Kernel-based method & \textbf{1.261}& \textbf{4.769}   \\
\hline
BRC & XGBoost & 3.940 & -43.79
   \\
 & Random Forest & 16.93 & -76.33   \\
 & Kernel-based method & \textbf{-0.5269}& \textbf{-14.40} \\
\hline
Max-call & XGBoost & -7.808 & -4.507   \\
 & Random Forest & -24.67 & -21.31   \\
 & Kernel-based method & \textbf{-0.3460}& \textbf{-3.588}   \\
\hline
\end{tabular}
\caption{Relative errors of value at risk $\mathrm{ES}_{99\%}(\mathrm{L}_{\bm X})$ and $\mathrm{ES}_{99\%}(-\mathrm{L}_{\bm X})$, computed as $(\text{estimated ES minus true ES})/\text{true ES}$ using the test sample and expressed in \%, using XGBoost and Random Forest, for the payoff functions min-put, BRC, and max-call.}\label{es_table}
\end{table}

%%% figure for minput

%%% figure for BRC

%%% figure for maxcall

\section{Outlook}\label{sec_outlook}
In this section we discuss two future research directions. The first one is computational: how to deal with higher dimensional problems? The second one is theoretical: is the value process estimator $V_{\bm X}$ asymptotically consistent?
\subsection{Scalability}\label{sec_scalability}
To apply our ensemble learning method to high dimensional problems, where both the sample size $n$ and the path space dimension $d\times T$ are very large, two conditions must be satisfied. First, fast training of the ensemble estimator $f_{\bm X}$ in \eqref{generic_el_estimator}. Second, fast evaluation of the value process estimator $V_{\bm X}$ in \eqref{eqYhat}.

Fortunately, the first condition is already satisfied. As shown in Table \ref{table_training_time_fx}, the training of $f_{\bm X}$ is extremely fast. This speed comes from two sources: a relatively small training time complexity, and the exploitation of parallel processing. In fact, the training time complexity for building a Random Forest $f_{\bm X, \bm \Pi}$ in \eqref{rand_forest}, under the bootstrapping sampling regime, is $\Theta(Mp  \tilde{n}\log(\tilde{n}))$, where $\tilde{n}=0.632 n$, see \cite[Table~5.1]{lou_2014}. And if we consider the XGBoost implementation of Gradient Boosting $f_{\bm X, t}$ in \eqref{gradient_descent}, this complexity becomes $\Theta(M\textbf{max\_depth}dT n\log(n))$, see \cite[Section~Time Complexity Analysis]{che_gue_2016}. The notation $\Theta(g(n))$, for some function $g$, means that there exist constants $0<c<C$, such that the training time complexity is bounded from below by $cg(n)$, and bounded from above by $Cg(n)$, as $n\to \infty$. Furthermore, the Random Forest implementation RandomForestRegressor in scikit-learn \cite{scikit_learn}, and the Gradient Boosting implementation XGBRegressor in XGBoost \cite{che_gue_2016} take advantage of parallel processing.

Now we shall discuss the second condition. The estimator $f_{\bm X}$ provided by a machine learning library, such as scikit-learn or XGBoost, should be rewritten into a suitable format. We recall the expression of $f_{\bm X}$ in \eqref{generic_el_estimator}. Thus $f_{\bm X}$ is determined by the hyperrectangles $\bm A_1, \dots,\bm  A_N$, and the real coefficients $\beta_1, \dots, \beta_N$. And from \eqref{box_notation}, we know that every hyperrectangle $\bm A_i$ is characterized by two matrices $a^{(i)}, b^{(i)} \in \R^{d\times T}$ such that $\bm A_i=( a^{(i)}, b^{(i)}]$. In summary, $f_{\bm X}$ is determined by the tuple $(\mathbf{cells}, \mathbf{values})$, where $\mathbf{cells} = (( a^{(1)}, b^{(1)}), \dots, (a^{(N)},b^{(N)}))$, and $\mathbf{values} = (\beta_1, \dots, \beta_N)$. The rewriting of $f_{\bm X}$ as $(\mathbf{cells}, \mathbf{values})$ forms a processing step, and it requires a careful look at how $f_{\bm X}$ is encoded in the corresponding library.\footnote{Python codes corresponding to the processing steps of RandomForestRegressor of scikit-learn, and XGBRegressor of XGBoost are available from the authors upon request.} After this processing step, the pseudo-code in Algorithm \ref{algo_eval_Vxt} shows how to evaluate $V_{\bm X, t}$ at some point $(x_1, \dots, x_t) \in \R^{d\times t}$. Thus fast evaluation of the value process $V_{\bm X}$ can be achieved by writing a fast code of Algorithm \ref{algo_eval_Vxt}.

Recall that in Section \ref{secexamples}, for each payoff function, we evaluate the function $V_{\bm X, 1}$ on a test sample of size $n_{\mathrm{test}}=100{,}000$. To achieve this we parallelized Algorithm \ref{algo_eval_Vxt} using the packages MPI for Python \cite{mpi4py} and joblib \cite{joblib}. Specifically, MPI for Python allowed us to parallelize the $100{,}000$ evaluations using several compute nodes. And joblib allowed us to parallelize the for loop in Algorithm \ref{algo_eval_Vxt}. Table \ref{table_time_vx} shows the time to evaluate $V_{\bm X, 1}$. We observe that evaluations of $V_{\bm X, 1}$ are faster with XGBoost than with Random Forest, which is mainly due to the difference in the number of hyperrectangles $N$ in these two estimators. The time to evaluate $V_{\bm X, 1}$ in Table \ref{table_time_vx} can be shortened further using, e.g., a GPU code of Algorithm \ref{algo_eval_Vxt}.

\begin{algorithm}[H]
\SetAlgoLined
\textbf{Input: $\mathbf{cells}$, $\mathbf{values}$, $t$, $(x_1, \dots, x_t)$}\;
{$\mathrm{value} = 0$}\;
{$\mathrm{ncells} = \mathrm{length}(\mathbf{values})$}\;
\For{$i=1, \dots, \mathrm{ncells}$}{
$(a,b) = \mathbf{cells}[i]$\;
\If{$a_{j,s}< x_{j,s}\le b_{j,s}$,  $j=1,\dots, d,$ and $s=1,\dots,t$}{
$\mathrm{value} = \mathrm{value} + \mathbf{values}[i] \times \prod_{s=t+1}^T\Q_s[(a_{s},b_{s}]]$
}}{
\textbf{return} $\mathrm{value}$\;
}
 \caption{Evaluation of $V_{\bm X, t}$}\label{algo_eval_Vxt}
\end{algorithm}

\begin{table}
\centering
  \begin{tabular}{|l|l|l|}
\hline
Payoff & Estimator  & Training time (s)     \\
\hline
Min-put & XGBoost  & 2   \\
 & Random Forest&  1 \\
\hline
BRC &XGBoost  & 8   \\
 & Random Forest& 12  \\
\hline
Max-call &XGBoost  & 2  \\
 & Random Forest&1   \\
\hline
\end{tabular}
\caption{Training time, in seconds, of $f_{\bm X}$, using XGBoost and Random Forest, using the training sample $\bm X$ and the function values $\bm f$. Computation is performed on 5 compute nodes, each has 2 Skylake processors running at 2.3 GHz, with 18 cores per processor. And we used 188 GB of RAM. We needed a large amount of RAM in order to evaluate $V_1$ and $V_{\bm X,1}$ on a test sample of size $n_{\mathrm{test}}=100{,}000$, see Section \ref{secexamples}. However, for the training of $f_{\bm X}$ one just needs a sufficient amount of memory to store $\bm X$ and $\bm f$.}\label{table_training_time_fx}
\end{table}

\begin{table}
\centering
  \begin{tabular}{|l|l|l|l|}
\hline
Payoff & Estimator  & Number of hyperrectangles $N$ & Time to evaluate $V_{\bm X, 1}$ (s)     \\
\hline
Min-put & XGBoost& 88{,}127 & 413   \\
 & Random Forest&494{,}118  &1{,}629 \\
\hline
BRC &XGBoost&189{,}864  & 10{,}947   \\
 & Random Forest&187{,}710 &13{,}472  \\
\hline
Max-call &XGBoost& 66{,}225 & 353  \\
 & Random Forest&489{,}747 &1{,}644   \\
\hline
\end{tabular}
\caption{Number of hyperrectangles $N$ in the ensemble estimator $f_{\bm X}$ in \eqref{generic_el_estimator}, using XGBoost and Random Forest, and time in seconds to evaluate $V_{\bm X,1}$ on a test sample of size $n_{\mathrm{test}}=100{,}000$. The number of hyperrectangles are copied from Tables \ref{table_parameters_rf} and \ref{table_parameters_xgb}. Computation is performed on 5 compute nodes, each has 2 Skylake processors running at 2.3 GHz, with 18 cores per processor. And we used 188 GB of RAM. We needed a large amount of RAM in order to evaluate $V_1$ and $V_{\bm X,1}$ on a test sample of size $n_{\mathrm{test}}=100{,}000$, see Section \ref{secexamples}. However, for the training of $f_{\bm X}$ one just needs a sufficient amount of memory to store $\bm X$ and $\bm f$.}\label{table_time_vx}
\end{table}

\subsection{Consistency}\label{sec_theory}

In Section \ref{secexamples}, we saw that our estimator $V_{\bm X}$ in \eqref{eqYhat} gives an accurate estimation of the value process $V$ in \eqref{eqcondY_EU}. In order to theoretically characterize the goodness of the estimator $V_{\bm X}$, from Doob’s maximal inequality in \eqref{doobineq}, we see that it is enough to study the ensemble estimator $f_{\bm X}$. In particular, we would be interested in consistency results of the form $\E_{\bm{Q}}[ \|f-f_{\bm X}\|^2_{2,\Q}]\to 0$, as $n\to \infty$, and finite sample guarantees of the form $\bm Q[\|f-f_{\bm X}\|^2_{2,\Q}<c(\eta, n)]\ge 1-\eta$, for all $n\ge n_0(\eta)$, for $\eta\in (0, 1]$. However, theoretical analysis of Random Forest, when $f_{\bm X}$ is $f_{\bm X, \bm \Pi}$ in \eqref{rand_forest}, and Gradient Boosting, when $f_{\bm X}$ is $f_{\bm X, t}$ in  \eqref{gradient_descent}, is difficult in general, especially in our financial framework, where the function $f$ typically is neither bounded nor compactly supported, see example \eqref{intro_example}. Below we mention two recent consistency results available in the machine learning literature that drew our attention and that could be investigated further.

Despite the plethora of empirical works on Random Forest, see, e.g., \cite{gen_et_al_2008}, \cite{arc_kim_2008}, and \cite{gen_et_al_2010}, to name a few, little is known on the theoretical side. The sampling scheme, the split criterion \eqref{fct_to_max}, and the sampling of $p$ coordinates for each hyperrectangle to split make the trees highly and non-trivially $(\bm X, \bm f)$-dependent. This renders the Random Forest difficult to analyse mathematically. To better understand theoretically the good performance of Random Forest in practice, less $(\bm X, \bm f)$-dependent versions of Random Forest have been studied, e.g., $(\bm X, \bm f)$-independent in \cite{bia_2012}, or $\bm f$-independent in \cite{sco_2016}. A recent consistency result for the asymptotic Random Forest, $\lim_{M \to \infty} f_{\bm X, \bm \Pi}$ in \eqref{rand_forest}, has been given in \cite{sco_et_al_2015}. They show that $\E_{\bm Q}[\|\lim_{M\to \infty}f_{\bm X, \bm \Pi} - f\|_{2,\Q}^2] \to 0 $, as $n\to \infty$, under the following assumptions: the path space is the unit cube $[0,1]^{d\times T}$ instead of $\R^{d\times T}$, $X$ is uniformly distributed on $[0,1]^{d\times T}$, and $f$ is continuous and additive. The last assumption reads in our case that $f(x) = \sum_{t=1}^{T} \sum_{j=1}^d f_{j, t}(x_{j, t})$, where each $f_{j, t}$ is continuous. These assumptions are too stringent in applications in finance. We recommend the survey \cite{bia_sco_2016} for an overview on the theoretical work on Random Forest.

As for Gradient Boosting, to the best of our knowledge, there is no consistency result for Gradient Boosting with CART in the context of regression problems in the literature. Nevertheless, we shall mention the recent paper \cite{bia_cad_2017}, where the authors study Gradient Boosting in both classification and regression problems. Their result \cite[Theorem~4.1]{bia_cad_2017} holds in the case where the base estimator is a certain type of regression trees. However, it does not hold in the case where the base estimator is a CART regression tree.

% , see, e.g., Blanchard et al. \cite{bla_et_al_2003}, Mannor et al. \cite{man_et_al_2003}, Lugosi and Vayatis \cite{lug_vay_2004}, Zhang and Yu \cite{zha_yu_2005}, Bickel et al. \cite{bic_et_al_2006}, Bartlett and Traskin \cite{bar_tar_2007}. 

\section{Conclusion}\label{secconc}

We introduce a unified framework for quantitative portfolio risk management based on the dynamic value process of the portfolio. We use ensemble estimators with regression trees to learn the value process from a finite sample of the cumulative cash flow of the portfolio. Our portfolio value process estimator is fast to construct, given in closed form, and accurate. The last means that the normalized $L^2_\Q$-error $\|\max_{t=0,\dots,T} | V_t-V_{\bm X,t}|\|_{2,\Q}/V_0$ is relatively small. In fact, numerical experiments for exotic and path-dependent options in the multivariate Black–Scholes model in moderate dimensions show good results for a moderate training sample size. In contrast to the kernel-based method in \cite{bou_fil_22}, our ensemble learning method can be scaled to deal with high dimensional problems.

% \newpage
\begin{appendix}
% \clearpage

\section{Comparison with regress-now}\label{sec_regnow_reglat}

The method we develop in this paper gives an estimation of the entire value process $V$. In practice, one could be interested in the estimation of the portfolio value $V_t$ only at some fixed time $t$, e.g., $t=1$. In \cite{gla_yu_2004} two least squares Monte Carlo methods are described to deal with this problem in the context of American options pricing. Their first method, termed ``regress-later'', consists in estimating the payoff function $f$ by means of a projection on a finite number of basis functions. The basis functions are chosen such that their conditional expectation at time $t=1$ is in closed form. Our method can be seen as a double extension of this, because it covers the case where both the basis functions and their number are not known a priori, and it gives closed-form estimation of the portfolio value $V_t$ at any time $t$. Their second method, termed ``regress-now'', consists in estimating $V_1$ by means of a projection on a finite number of basis functions that depend solely on the variable of interest $x_1 \in \R^d$.

We compare our method, which corresponds to ``regress-later'', and which gives the estimator $V_{\bm X, t}$ in \eqref{hat_Vt_closed_form} for $t=1$, to its regress-now variant, whose estimator we denote by $V_{\bm X, 1}^{\text{now}}$. Thereto we briefly discuss how to construct $V_{\bm X, 1}^{\text{now}}$ in the context of the three payoff functions studied in Section \ref{secexamples}.

The construction of $V_{\bm X, 1}^{\text{now}}$ is simpler than that of $V_{\bm X, 1}$. First, instead of the whole sample $\bm X$, one only needs the $(t=1)$-cross-section $\bm X_1 = (X^{(1)}_1,\dots, X^{(n)}_1)$. Second, with the input $\bm X_1$ and $\bm f$, the estimators Random Forest in Section \ref{sec_rf}, and Gradient Boosting in Section \ref{sec_gb} give directly $V_{\bm X, 1}^{\text{now}}$. As we did in Section \ref{secexamples}, we use the validation sample $\bm X_{\mathrm{valid}}$, along with its corresponding function values $\bm f_{\mathrm{valid}}$, to find the optimal hyperparameter values for Random Forest and Gradient Boosting by validation on the sets of hyperparameter values $\Pcal_{\mathrm{RF}}$ and $\Pcal_{\mathrm{XGB}}$, respectively. Table \ref{table_parameters_rf_now} shows the normalized $L^2_\Q$-error $\|V_{\bm X, 1}^{\mathrm{now}}-f\|_{2,\Q}/V_0$, computed using the validation sample $\bm X_{\mathrm{valid}}$, and the number of hyperrectangles $N$ in the Random Forest and Gradient Boosting $f_{\bm X}$ in \eqref{generic_el_estimator} for the optimal hyperparameter values in $\Pcal_{\mathrm{RF}}$ and $\Pcal_{\mathrm{XGB}}$. Unlike in Section \ref{secexamples}, here we choose the optimal hyperparameter value for Random Forest, although the number of hyperrectangles $N$ induced is very large ($N> 500{,}000$). This is because the evaluation of $V_{\bm X, 1}^{\mathrm{now}}$ is very fast irrespectively of $N$, thanks to the highly optimized implementations of Random Forest and Gradient Boosting, RandomForestRegressor in scikit-learn \cite{scikit_learn} and XGBRegressor in XGBoost \cite{che_gue_2016}, respectively.

\begin{table}
\centering
  \begin{tabular}{|l|l|l|l|}
\hline
 & Min-put  & BRC & Max-call     \\
\hline
Optimal hyperparameter value for Random Forest & (500, 5, 2)  & (500, 5, 1) &(500, 5, 2)   \\
Normalized $L^2_\Q$-error in \% &31.98 & 7.195 & 46.58  \\
Number of hyperrectangles &2{,}619{,}900 & 323{,}896 & 2{,}620{,}048 \\
\hline
Optimal hyperparameter value for Gradient Boosting & (44, 40, 45)  & (52, 40, 45) &(41, 50, 45) \\
Normalized $L^2_\Q$-error in \% & 32.43 & 6.886 & 46.62 \\
Number of hyperrectangles &13{,}401 & 9{,}606 & 13{,}165 \\
\hline
\end{tabular}
\caption{
Regress-now Random Forest and regress-now XGBoost validation steps: normalized $L^2_\Q$-error $\|V_{\bm X,1}^{\mathrm{now}}-f\|_{2, \Q}/V_0$, computed using the validation sample $\bm X_{\mathrm{valid}}$ and expressed in \%, and number of hyperrectangles $N$ in the Random Forest and Gradient Boosting $f_{\bm X}$ in \eqref{generic_el_estimator}, for the optimal hyperparameter values in $\Pcal_{\mathrm{RF}}$ and $\Pcal_{\mathrm{XGB}}$, for the payoff functions min-put, BRC, and max-call.}\label{table_parameters_rf_now}
\end{table}

\begin{figure}[p]
    \centering % <-- added
    \begin{subfigure}{0.45\textwidth}
  \includegraphics[width=\linewidth]{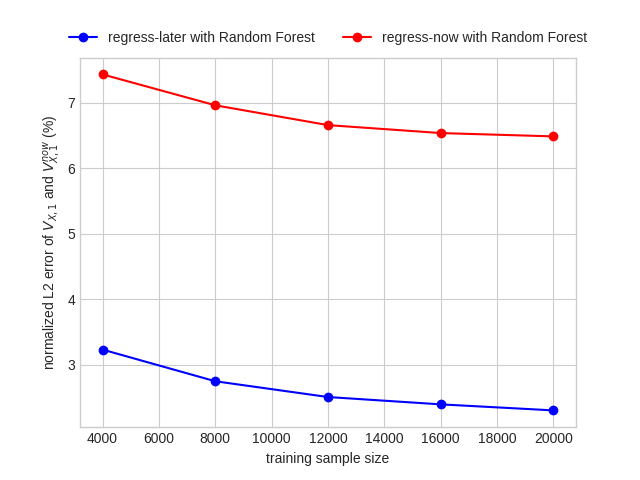}
  \caption{Min-put with Random Forest: normalized $L^2_{\Q}$-errors of $V_{\bm X, 1}$ and $V_{\bm X, 1}^{\text{now}}$ in \%.}
  \label{error_V1_minput_rho0_nowVSlater_rf}
\end{subfigure}\hfil % <-- added
\begin{subfigure}{0.45\textwidth}
  \includegraphics[width=\linewidth]{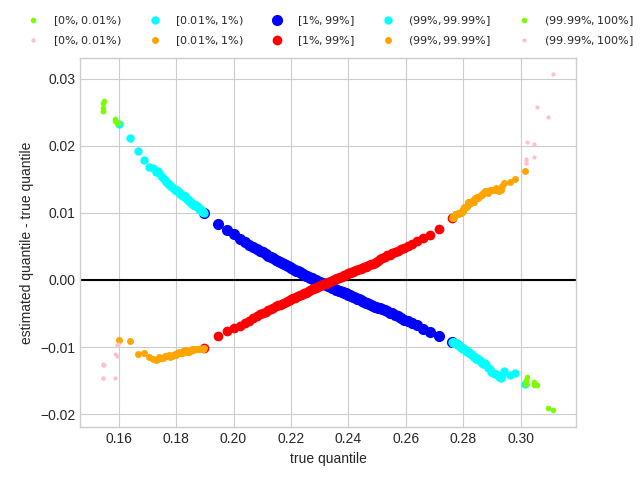}
  \caption{Min-put with Random Forest: detrended Q-Q plots of $V_{\bm X, 1}$ and $V_{\bm X, 1}^{\text{now}}$.}
  \label{qqplot_minput_rho0_nowVSlater_rf}
\end{subfigure}
\medskip
\begin{subfigure}{0.45\textwidth}
  \includegraphics[width=\linewidth]{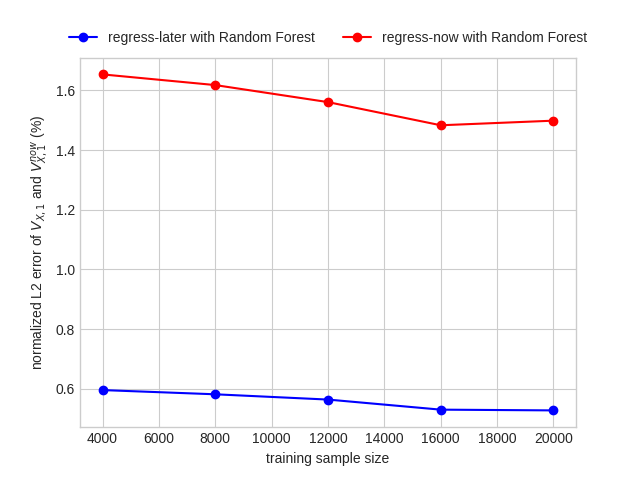}
  \caption{BRC with Random Forest: normalized $L^2_{\Q}$-errors of $V_{\bm X, 1}$ and $V_{\bm X, 1}^{\text{now}}$ in \%.}
  \label{error_V1_barrier_rho0_nowVSlater_rf}
\end{subfigure}\hfil % <-- added
\begin{subfigure}{0.45\textwidth}
  \includegraphics[width=\linewidth]{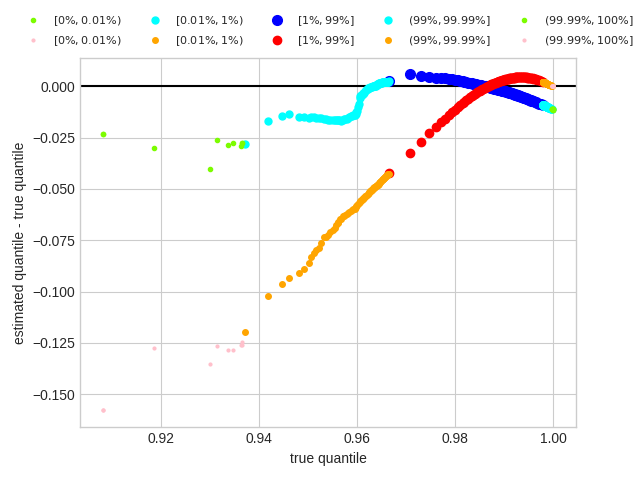}
  \caption{BRC with Random Forest: detrended Q-Q plots of $V_{\bm X, 1}$ and $V_{\bm X, 1}^{\text{now}}$.}
  \label{qqplot_barrier_rho0_nowVSlater_rf}
\end{subfigure}
\medskip
\begin{subfigure}{0.45\textwidth}
  \includegraphics[width=\linewidth]{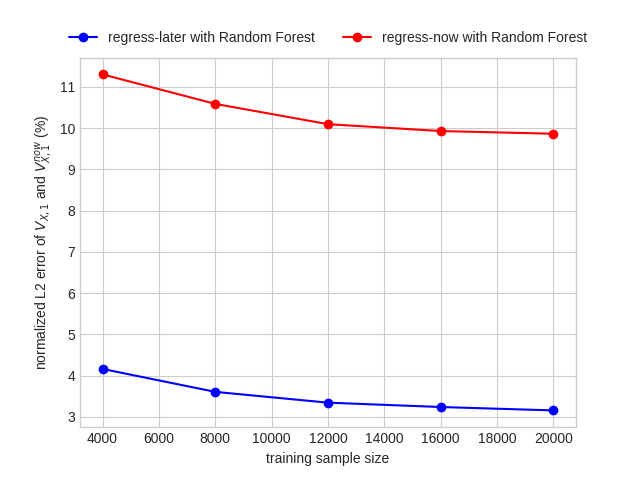}
  \caption{Max-call with Random Forest: normalized $L^2_{\Q}$-errors of $V_{\bm X, 1}$ and $V_{\bm X, 1}^{\text{now}}$ in \%.}
  \label{error_V1_maxcall_rho0_nowVSlater_rf}
\end{subfigure}\hfil % <-- added
\begin{subfigure}{0.45\textwidth}
  \includegraphics[width=\linewidth]{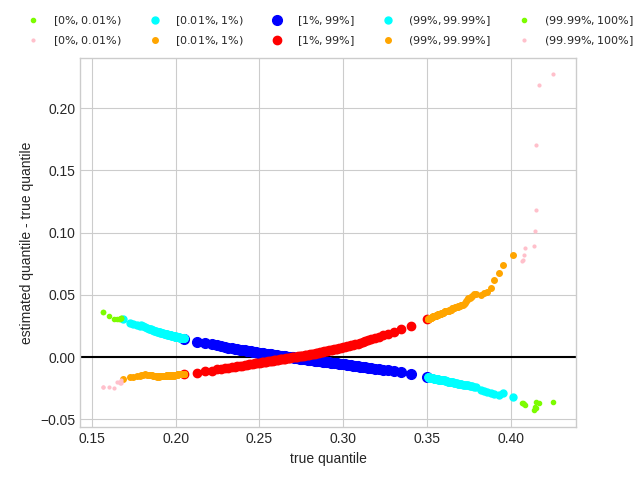}
  \caption{Max-call with Random Forest: detrended Q-Q plots of $V_{\bm X, 1}$ and $V_{\bm X, 1}^{\text{now}}$.}
  \label{qqplot_maxcall_rho0_nowVSlater_rf}
\end{subfigure}
\caption{
Results for the min-put, BRC, and max-call with Random Forest. The normalized $L^2_\Q$-errors of $V_{\bm X, 1}$ and $V_{\bm X, 1}^{\text{now}}$, $\|V_1-V_{\bm X, 1}\|_{2, \Q}/V_0$ and $\|V_1-V_{\bm X, 1}^{\text{now}}\|_{2, \Q}/V_0$, are computed using the test sample and expressed in \%. In the detrended Q-Q plots, the blue, cyan, and lawngreen (red, orange, and pink) dots are built using regress-later Random Forest (regress-now Random Forest) and the test sample. $[0\%, 0.01\%)$ refers to the quantiles of levels $\{0.001\%,0.002\%, \dots, 0.009\%\}$, $[0.01\%, 1\%)$ refers to the quantiles of levels $\{0.01\%,0.02\%, \dots, 0.99\%\}$, $[1\%, 99\%]$ refers to the quantiles of levels $\{1\%,2\%, \dots, 99\%\}$, $(99\%, 99.99\%]$ refers to the quantiles of levels $\{99.01\%,99.02\%, \dots, 99.99\%\}$, and $(99.99\%, 100\%]$ refers to the quantiles of levels $\{99.991\%,99.992\%, \dots, 100\%\}$.}
\label{fig_nowVSlater_rf}
\end{figure}

\begin{figure}[p]
    \centering % <-- added
    \begin{subfigure}{0.45\textwidth}
  \includegraphics[width=\linewidth]{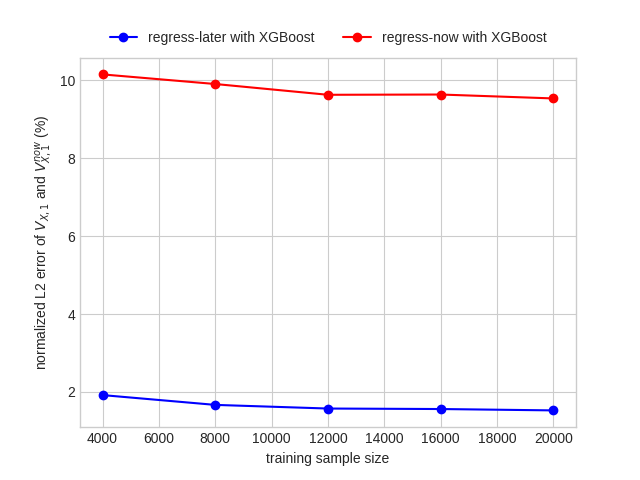}
  \caption{Min-put with XGBoost: normalized $L^2_{\Q}$-errors of $V_{\bm X, 1}$ and $V_{\bm X, 1}^{\text{now}}$ in \%.}
  \label{error_V1_minput_rho0_nowVSlater_xgb}
\end{subfigure}\hfil % <-- added
\begin{subfigure}{0.45\textwidth}
  \includegraphics[width=\linewidth]{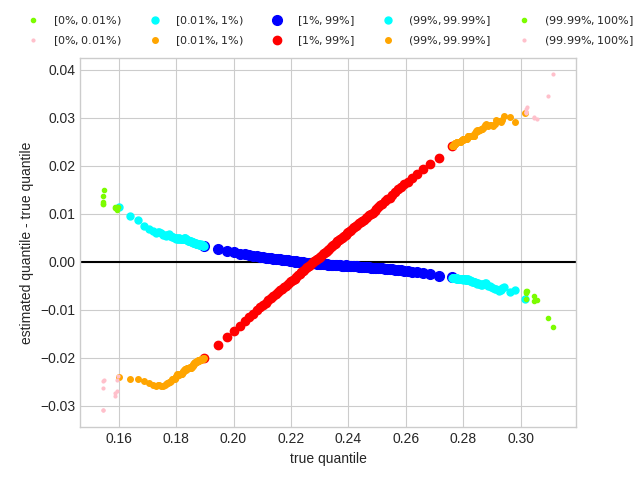}
  \caption{Min-put with XGBoost: detrended Q-Q plots of $V_{\bm X, 1}$ and $V_{\bm X, 1}^{\text{now}}$.}
  \label{qqplot_minput_rho0_nowVSlater_xgb}
\end{subfigure}
\medskip
\begin{subfigure}{0.45\textwidth}
  \includegraphics[width=\linewidth]{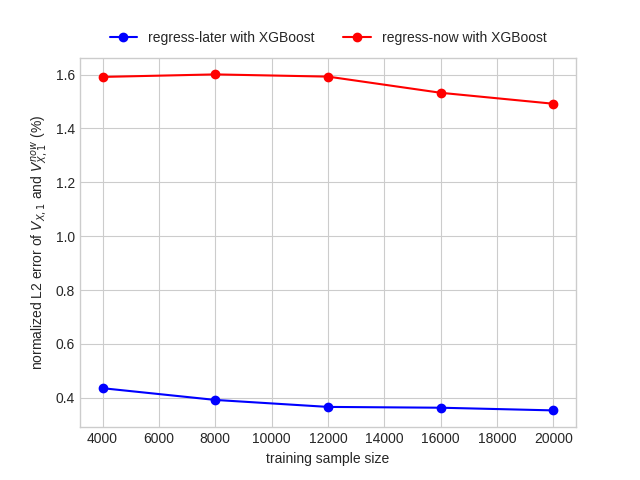}
  \caption{BRC with XGBoost: normalized $L^2_{\Q}$-errors of $V_{\bm X, 1}$ and $V_{\bm X, 1}^{\text{now}}$ in \%.}
  \label{error_V1_barrier_rho0_nowVSlater_xgb}
\end{subfigure}\hfil % <-- added
\begin{subfigure}{0.45\textwidth}
  \includegraphics[width=\linewidth]{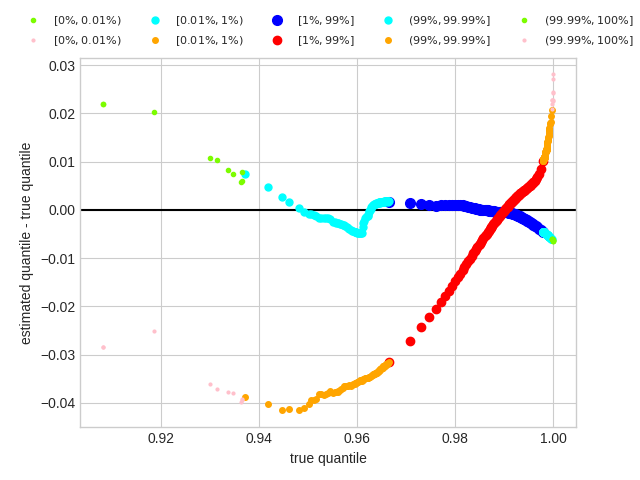}
  \caption{BRC with XGBoost: detrended Q-Q plots of $V_{\bm X, 1}$ and $V_{\bm X, 1}^{\text{now}}$.}
  \label{qqplot_barrier_rho0_nowVSlater_xgb}
\end{subfigure}
\medskip
\begin{subfigure}{0.45\textwidth}
  \includegraphics[width=\linewidth]{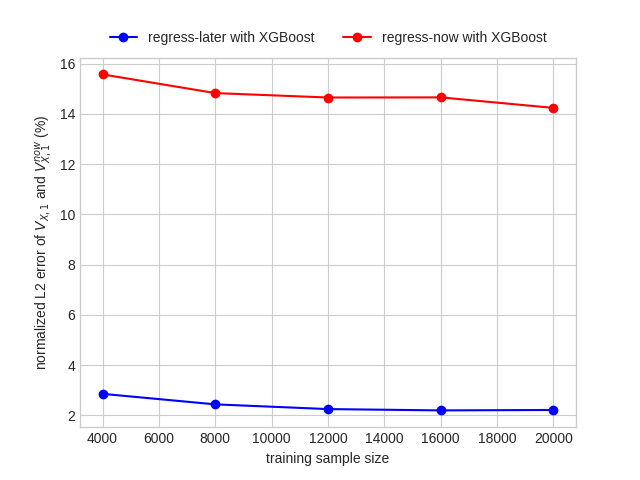}
  \caption{Max-call with XGBoost: normalized $L^2_{\Q}$-errors of $V_{\bm X, 1}$ and $V_{\bm X, 1}^{\text{now}}$ in \%.}
  \label{error_V1_maxcall_rho0_nowVSlater_xgb}
\end{subfigure}\hfil % <-- added
\begin{subfigure}{0.45\textwidth}
  \includegraphics[width=\linewidth]{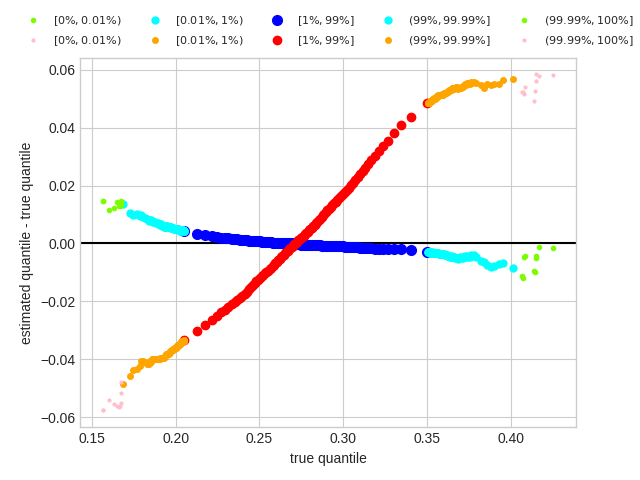}
  \caption{Max-call with XGBoost: detrended Q-Q plots of $V_{\bm X, 1}$ and $V_{\bm X, 1}^{\text{now}}$.}
  \label{qqplot_maxcall_rho0_nowVSlater_xgb}
\end{subfigure}
\caption{Results for the min-put, BRC, and max-call with XGBoost. The normalized $L^2_\Q$-errors of $V_{\bm X, 1}$ and $V_{\bm X, 1}^{\text{now}}$, $\|V_1-V_{\bm X, 1}\|_{2, \Q}/V_0$ and $\|V_1-V_{\bm X, 1}^{\text{now}}\|_{2, \Q}/V_0$, are computed using the test sample and expressed in \%. In the detrended Q-Q plots, the blue, cyan, and lawngreen (red, orange, and pink) dots are built using regress-later XGBoost (regress-now XGBoost) and the test sample. $[0\%, 0.01\%)$ refers to the quantiles of levels $\{0.001\%,0.002\%, \dots, 0.009\%\}$, $[0.01\%, 1\%)$ refers to the quantiles of levels $\{0.01\%,0.02\%, \dots, 0.99\%\}$, $[1\%, 99\%]$ refers to the quantiles of levels $\{1\%,2\%, \dots, 99\%\}$, $(99\%, 99.99\%]$ refers to the quantiles of levels $\{99.01\%,99.02\%, \dots, 99.99\%\}$, and $(99.99\%, 100\%]$ refers to the quantiles of levels $\{99.991\%,99.992\%, \dots, 100\%\}$.}
\label{fig_nowVSlater_xgb}
\end{figure}

Table \ref{norm_l2_errors_nowVSlater} shows the normalized $L^2_{\Q}$-errors of $V_{\bm X, 1}$ and $V_{\bm X, 1}^{\text{now}}$. The former values are copied from Table \ref{norm_l2_errors_table} for convenience. We observe that our regress-later estimators always perform better than their regress-now variants. This finding is confirmed by the normalized $L^2_\Q$-errors of $V_{\bm X, 1}$ and $V_{\bm X, 1}^{\text{now}}$ as function of the training sample in Figures \ref{error_V1_minput_rho0_nowVSlater_rf}, \ref{error_V1_barrier_rho0_nowVSlater_rf}, \ref{error_V1_maxcall_rho0_nowVSlater_rf}, \ref{error_V1_minput_rho0_nowVSlater_xgb}, \ref{error_V1_barrier_rho0_nowVSlater_xgb}, \ref{error_V1_maxcall_rho0_nowVSlater_xgb}. In \cite{bou_fil_22} we also compare regress-later and regress-now with the kernel-based method. In Table \ref{norm_l2_errors_nowVSlater} we report, from \cite[Table~5]{bou_fil_22}, the $L^2_\Q$-errors corresponding to the kernel-based method. We see that also with the kernel-based method, regress-later outperforms regress-now.

Now we discuss the detrended Q-Q plots in Figures \ref{qqplot_minput_rho0_nowVSlater_rf}, \ref{qqplot_barrier_rho0_nowVSlater_rf}, \ref{qqplot_maxcall_rho0_nowVSlater_rf} for Random Forest, and in Figures \ref{qqplot_minput_rho0_nowVSlater_xgb}, \ref{qqplot_barrier_rho0_nowVSlater_xgb}, \ref{qqplot_maxcall_rho0_nowVSlater_xgb} for XGBoost. Their construction is detailed in the second-to-last paragraph of Section \ref{secexamples}. In Figures \ref{qqplot_minput_rho0_nowVSlater_xgb}, \ref{qqplot_barrier_rho0_nowVSlater_xgb}, \ref{qqplot_maxcall_rho0_nowVSlater_xgb}, we see that for XGBoost, the detrended Q-Q plots with regress-later are of much better quality, i.e., they are more aligned with the horizontal black line, than the detrended Q-Q plots with regress-now. As for Random Forest, the outperformance of regress-later over regress-now in terms of normalized $L^2_\Q$-error, seen in Figures \ref{error_V1_minput_rho0_nowVSlater_rf}, \ref{error_V1_barrier_rho0_nowVSlater_rf}, \ref{error_V1_maxcall_rho0_nowVSlater_rf}, does not clearly appear in the detrended Q-Q plots in Figures \ref{qqplot_minput_rho0_nowVSlater_rf}, \ref{qqplot_barrier_rho0_nowVSlater_rf}, \ref{qqplot_maxcall_rho0_nowVSlater_rf}. By comparing Figures \ref{qqplot_minput_rho0_nowVSlater_rf} and \ref{qqplot_minput_rho0_nowVSlater_xgb}; Figures \ref{qqplot_barrier_rho0_nowVSlater_rf} and \ref{qqplot_barrier_rho0_nowVSlater_xgb}; Figures \ref{qqplot_maxcall_rho0_nowVSlater_rf} and \ref{qqplot_maxcall_rho0_nowVSlater_xgb}, we see that regress-later XGBoost gives the best detrended Q-Q plots, i.e., the detrended Q-Q plots that are the most aligned with the horizontal black line.

We finish this section by discussing the risk measure estimates in Tables \ref{var_table_nowVlater} and \ref{es_table_nowVlater}. Their construction is detailed in the last paragraph of Section \ref{secexamples}. Consistently with the last comment in the above paragraph, it is regress-later XGBoost that gives best estimates of risk measures in most cases, 10 cases out of 12. In the 2 left cases, which correspond to the estimation of risk measures of the short position of the BRC, it is regress-now Random Forest that is the most accurate. The last is consistent with the detrended Q-Q plots in Figures \ref{qqplot_barrier_rho0_nowVSlater_rf} and \ref{qqplot_barrier_rho0_nowVSlater_xgb}, where we observe that regress-now Random Forest gives the best estimation of the right tail distribution of $V_1$ among all estimators.

%%% normalized L2 errors table for regress-now and regress-later
\begin{table}
\centering
  \begin{tabular}{|l|l|l|l|}
\hline
Payoff &  Estimator &         $V_{\bm X, 1}$ &         $V_{\bm X, 1}^{\text{now}}$ \\
\hline
Min-put & XGBoost & \underline{\bf{1.525}}&   9.539 \\
 & Random Forest  & \bf{2.300} &  6.487  \\
  & Kernel-based method  & \textbf{1.827}& 1.946  \\
\hline
BRC & XGBoost  & \bf{0.3530} & 1.492   \\
 &  Random Forest &  \bf{0.5276} & 1.499   \\
     & Kernel-based method  & \underline{\textbf{0.2506}}& 0.2806   \\
\hline
Max-call & XGBoost & \underline{\bf{2.217}}&  14.24  \\
 & Random Forest &\bf{3.155} &   9.863 \\
 & Kernel-based method  & \textbf{2.315}& 2.606   \\
\hline
\end{tabular}
\caption{
Normalized $L^2_\Q$-error $\|V_1  -  \widehat{V}_{1} \|_{2, \Q}/V_0$, computed using the test sample and expressed in \%, for $\widehat{V}_{1}\in \{V_{\bm X, 1}, V_{\bm X, 1}^{\text{now}}\}$, for the payoff functions min-put, BRC, and max-call.}\label{norm_l2_errors_nowVSlater}
\end{table}

%%% VAR table regnow vs reglatter
\begin{table}
\centering
  \begin{tabular}{|l|l|l|l|l|l|}
\hline
Payoff & Estimator &  $\mathrm{VaR}(\mathrm{L}_{\bm X})$& $\mathrm{VaR}(\mathrm{L}_{\bm X}^{\text{now}})$ & $\mathrm{VaR}(-\mathrm{L}_{\bm X})$ & $\mathrm{VaR}(-\mathrm{L}_{\bm X}^{\text{now}})$\\
\hline
Min-put & XGBoost & \textbf{-9.658} & 50.94 & \textbf{-6.912} & 48.21   \\
 & Random Forest & -25.69 & \textbf{21.38} & \textbf{-20.57} & 23.30  \\
  & Kernel-based method & \underline{\textbf{0.9695}}& 1.697& \underline{\textbf{3.158}}& -9.184  \\
\hline
BRC & XGBoost  & \textbf{2.533} & 123.3 & \textbf{-42.85} & 141.8   \\
 & Random Forest & \textbf{-3.510} & 197.6 & -75.87 & \textbf{24.91}   \\
   & Kernel-based method & \underline{\textbf{0.1893}}&
 -16.81&
 -13.91&
 \underline{\textbf{-7.342}} \\
\hline
Max-call & XGBoost & \textbf{-7.103} & 50.97 & \textbf{-4.140} & 57.89   \\
 & Random Forest & -23.87 & \textbf{21.88} & \textbf{-20.51} & 40.00   \\
   & Kernel-based method & \underline{\textbf{0.07143}}&
 5.357&
 -3.582&
\underline{\textbf{1.237}}  \\
 \hline
\end{tabular}
\caption{
Relative errors of value at risk $\mathrm{VaR}_{99.5\%}(\mathrm{L}_{\bm X})$, $\mathrm{VaR}_{99.5\%}(\mathrm{L}_{\bm X}^{\mathrm{now}})$, $\mathrm{VaR}_{99.5\%}(-\mathrm{L}_{\bm X})$, $\mathrm{VaR}_{99.5\%}(-\mathrm{L}_{\bm X}^{\mathrm{now}})$, computed as $(\text{estimated VaR minus true VaR})/\text{true VaR}$ using the test sample and expressed in \%, using XGBoost and Random Forest, for the payoff functions min-put, BRC, and max-call.}\label{var_table_nowVlater}
\end{table}

%%% ES table regnow vs reglatter
\begin{table}
\centering
  \begin{tabular}{|l|l|l|l|l|l|}
\hline
Payoff & Estimator &  $\mathrm{ES}(\mathrm{L}_{\bm X})$& $\mathrm{ES}(\mathrm{L}_{\bm X}^{\text{now}})$ & $\mathrm{ES}(-\mathrm{L}_{\bm X})$ & $\mathrm{ES}(-\mathrm{L}_{\bm X}^{\text{now}})$\\
\hline
Min-put & XGBoost & \textbf{-10.23} & 49.75 & \textbf{-7.434} & 47.22   \\
 & Random Forest & -26.41 & \textbf{20.46} & \textbf{-21.07} & 23.68   \\
 & Kernel-based method & \underline{\textbf{1.261}}& 1.775& \underline{\textbf{4.769}}& -8.546  \\
\hline
BRC & XGBoost & \textbf{3.940} & 113.5 & \textbf{-43.79} & 151.1  \\
 &  Random Forest & \textbf{16.93} & 203.1 & -76.33 & \textbf{22.27}  \\
 & Kernel-based method & \underline{\textbf{-0.5269}}&
 -18.58&
 -14.40&
 \underline{\textbf{-8.271}}  \\
\hline
Max-call & XGBoost & \textbf{-7.808} & 50.12 & \textbf{-4.507} & 55.32   \\
 & Random Forest & -24.67 & \textbf{20.90} & \textbf{-21.31} & 42.63   \\
 & Kernel-based method & \underline{\textbf{-0.3460}}&
 5.329&
 -3.588&
 \underline{\textbf{0.8112}}  \\
 \hline
\end{tabular}
\caption{Relative errors of value at risk $\mathrm{ES}_{99\%}(\mathrm{L}_{\bm X})$, $\mathrm{ES}_{99\%}(\mathrm{L}_{\bm X}^{\mathrm{now}})$, $\mathrm{ES}_{99\%}(-\mathrm{L}_{\bm X})$, $\mathrm{ES}_{99\%}(-\mathrm{L}_{\bm X}^{\mathrm{now}})$, computed as $(\text{estimated ES minus true ES})/\text{true ES}$ using the test sample and expressed in \%, using XGBoost and Random Forest, for the payoff functions min-put, BRC, and max-call.}\label{es_table_nowVlater}
\end{table}

%%% normalized errors for regnow vs reglatter with random forest

%%% normalized errors for regnow vs reglatter with xgboost

%\newpage
\section{Bermudan options pricing}\label{sec_optimal_stopping}

So far we showed that ensemble estimators with regression trees can be employed to learn a value process of the form \eqref{eqcondY_EU}. In this section, we sketch how these estimators can also be applied to deal with another important and difficult problem in finance, the pricing of Bermudan options under discrete-time local volatility models.\footnote{For the sake of presentation, we only consider discrete-time local volatility models. However, the method we detail below can be adapted to also deal with discrete-time stochastic volatility models.} A standard solution to this problem is to recursively estimate the continuation value, see, e.g., \cite{tsi_roy_1999}, and \cite{lon_sch_2001}. Recent solutions based on machine learning techniques include \cite{bec_et_al_2019}, who apply deep neural networks to learn the optimal stopping rule, and \cite{gou_et_al_2020}, who apply Gaussian Process Regression to learn the value function at every time step. We add to this literature by showing that ensemble estimators with regression trees give closed-form estimators of the entire value process of Bermudan options. 

Similarly to what we discussed in Appendix \ref{sec_regnow_reglat} for European style options pricing, we first apply regress-later and then regress-now.

\subsection{Regress-later}\label{sec_us_option_reglater}
Assume the stochastic driver $X=(X_1, \dots, X_T)$ has standard normal distribution on $\R^{d\times T}$, i.e., $\Q_t = \Ncal(0, I_{d})$ for every $t=1, \dots, T$. Let $m \in \N$, and $\alpha_t : \R^{m} \to \R^{m}$ and $\beta_t : \R^{m} \to \R^{m\times d}$ be measurable functions, for $t=0,\dots, T-1$. Let $(Z_t)_{0\le t \le T}$ be an $m$-dimensional stochastic process that represents the evolution of the log-price of $m$ underlying assets. We assume that it follows the following discrete-time local volatility model, 
\begin{equation}\label{lv_model}
\textstyle
\begin{cases}
Z_{t} &= \alpha_{t-1}(Z_{t-1}) + \beta_{t-1}(Z_{t-1}) X_{t},\quad t =1, \dots, T,\\
Z_{0} &= z_0,
\end{cases}
\end{equation}
where $z_0 \in \R^{m}$ is the initial log-price vector. We denote by $\mathbb G=(\Gcal_t)_{0\le t\le T}$ the natural filtration of $Z$, $\Gcal_t =\sigma(Z_s\mid 0\le s\le t)$. Let $g_t: \R^{m} \to \R$ be measurable functions such that $\E_{ \Q}[g_t(Z_t)^2]< \infty$, for $t=0, \dots, T$. Let $\Tcal_t$ be the set of $\mathbb G$-stopping times $\tau$ taking values in $\{t,t+1,\dots,T\}$, for $t=0,\dots, T$. We are interested in the following optimal stopping problem,
\begin{equation}\label{eqcondY_US_markov}
\textstyle
    V_t = \sup_{\tau\in\Tcal_t} \E_{\Q}[g_{\tau}( Z_{\tau})\mid  \Gcal_t], \quad t=0, \dots, T.\footnote{We look at Markovian problems, in the sense that $g_t$ depends only on the state variable $Z_t$. However, our method can be adapted to also deal with non Markovian problems, where $g_t$ could depend on the whole past trajectory $Z_0, \dots, Z_t$.}
\end{equation}
It is well known, see, e.g., \cite[Section~1]{pes_shi_2006}, problem \eqref{eqcondY_US_markov} can be solved by backward induction as follows. For time $t=T$, set $V_T=  g_T(Z_T)$. By induction, for any time $t= T-1,\dots,0$, define the continuation value $C_{t} = \E_{ \Q}[V_{t+1}(Z_{t+1}) \mid Z_{t} ]$, so that $V_{t} = \max(g_{t}(Z_{t}), C_{t})$. Then, an optimal stopping time $\tau^\star_t\in\Tcal_t$, for which $V_t = \E_{ \Q}[g_{\tau^\star_t}(Z_{\tau^\star_t})\mid  \Gcal_t]$, is given by $\tau^\star_t = \inf\{t\le s \le T \mid V_s= g_s(Z_s)\}$.

Our method to solve \eqref{eqcondY_US_markov} is based on a backward induction, where for each time $t=T-1,\dots, 0$ we estimate the value function $V_t$ by an ensemble estimator $V_{\bm X, t}$. Specifically, assume available a finite i.i.d.\ sample $\bm{ Z} = (Z^{(1)}, \dots, Z^{(n)})$ drawn from \eqref{eqcondY_US_markov}. And let $\bm{Z}_t = (Z^{(1)}_t, \dots, Z^{(n)}_t)$ be the $t$-cross-section sample of $\bm{Z}$, for $t=1, \dots, T$. Then proceed backward as follows:
\begin{enumerate}
    \item For time $t=T$: set $V_{\bm X, T} =  g_T(Z_T)$.
    \item For any time $t= T-1,\dots,0$: let $\widehat{V}_{\bm X,t+1}$ be an ensemble estimator of $V_{\bm X,t+1}$, obtained using the sample $\bm{Z}_{t+1}$, along with the function values $\bm{V}_{\bm X, t+1} = (V_{\bm X, t+1}(Z_{t+1}^{(1)}), \dots, V_{\bm X, t+1}(Z_{t+1}^{(n)}))$. Then, we claim that
    \begin{equation}\label{ct_markov}
    \textstyle
        C_{\bm X, t} = \E_{\Q}[\widehat{V}_{\bm X, t+1}(Z_{t+1})\mid Z_{t}] \text{ is in \textbf{closed form}}.
    \end{equation}
     Then set $V_{\bm X, t} = \max( g_{t}(Z_{t}), C_{\bm X, t})$, which is therefore also in closed form.
\end{enumerate}

Now let us explain why $C_{\bm X, t}$ in \eqref{ct_markov} is in closed form. The function $\widehat{V}_{\bm X, t+1}$ is an ensemble estimator, whose expression can be brought into the form \eqref{generic_el_estimator}, i.e., $\widehat{V}_{\bm X, t+1} = \sum_{i=1}^{N} \beta_i \1_{\bm A_i}$, for some real coefficients $\beta_i$ and some hyperrectangles $\bm A_i$ of $\R^m$. Subsequently, by independence of $Z_{t}$ and $X_{t+1}$, we readily obtain that
\[\textstyle C_{\bm X,t}( z_{t}) = \sum_{i=1}^N \beta_{i} \Q_{t+1}[\alpha_{t}( z_{t}) + \beta_{t}( z_{t}) X_{t+1} \in \bm A_i ], \quad  z_{t} \in \R^m.\]
Now observe that $\alpha_{t}( z_{t}) + \beta_{t}( z_{t}) X_{t+1}$ is normally distributed under $\Q_{t+1}$, with mean $\alpha_{t}( z_{t})$ and covariance matrix $\beta_{t}( z_{t}) \beta_{t}( z_{t})^\top$. Thanks to \cite{genz_2000}, $\Q_{t+1}[\alpha_{t}( z_{t}) + \beta_{t}( z_{t}) X_{t+1} \in \bm A_i ]$ is in closed form for any hyperrectangle $\bm A_i$. In fact, functions to integrate a multivariate normal distribution on a hyperrectangle are readily accessible on scientific languages, such as Python (see the mvn function in the sub-package stats of the library SciPy \cite{scipy}), and R (see the function pmvnorm in the mvtnorm package \cite{genz_et_al_mvtnorm}). This became possible thanks to the Fortran code of \cite{genz_web}. Matlab code to integrate a multivariate normal distribution on a hyperrectangle can also be found in \cite{genz_web}.

After the construction of the value process estimator $V_{\bm X}$, we define the optimal stopping time estimator 
\begin{equation}\label{opt_stop_rule_est}
    \tau^{\star}_{\bm X, t} = \inf\{t\le s \le T \mid V_{\bm X, s} = g_s(Z_s)\}.
\end{equation}

Now as a simple numerical example, let $S_t$ represent the nominal price of an underlying asset, whose dynamics is given in \eqref{bs_model} with $d=1$. We denote by $Z_t = \log(S_t)$ its log-price. Then $Z_t$ follows the dynamics
\[\textstyle\begin{cases}
Z_{t} &= Z_{t-1} + (r-\sigma^2/2)\Delta_t + \sigma \sqrt{\Delta_t} X_{t},\quad t =1, \dots, T,\\
Z_{0} &= z_0,
\end{cases}\]
which is of the form \eqref{lv_model}. We are interested in estimating the value process $V$ in \eqref{eqcondY_US_markov}, where the payoff function $g_t$ is 
\begin{itemize}
    \item Put $g_t(Z_t) = \e^{-r\sum_{s=1}^t \Delta_s} (K-\e^{Z_t})^+$.
\end{itemize}
Here we set the following parameter values, $z_0 = 0$, $r=0$, $\sigma=0.2$, $T=7$, $(\Delta_1, \dots, \Delta_T)=(1/T, \dots, 1/T)$, and $K=1$. Under this parameter specification, we generate a training sample $\bm X$ of size $n= 5{,}000$, and a test sample $\bm X_{\mathrm{test}}$ of size $n_{\mathrm{test}}=100{,}000$. When $V_{\bm X}$ is the Random Forest estimator of $V$, we use the RandomForestRegressor class of the library scikit-learn \cite{scikit_learn}, with the following hyperparameter values: $M$=10, \textbf{nodesize} = 2, $p=1$, \textbf{sampling
regime}=bootstrapping.\footnote{Except for the number of trees $M$, the other hyperparameter values are the default values in RandomForestRegressor. We picked a small value for $M$ for computational reasons.} When $V_{\bm X}$ is the Gradient Boosting estimator of $V$, we use the XGBRegressor class of XGBoost \cite{che_gue_2016}, with the default hyperparameter values: $t=100$, $\textbf{nodesize}=1$, $\textbf{max\_depth}=6$.

With the Bermudan put example and the parameter specification, $r=0$, one can readily show that early stopping is not optimal, so that $V_t=\E[(K-\e^{Z_{T}})^+\mid Z_t]$ equals the European put option price, and $V_t$ is given in closed form thanks to Black's formula. In fact, for $t=0, \dots, T-1$, $V_t = -\e^{Z_t} \Phi(-d_1) + K \Phi(-d_2)$, where $d_1= \frac{1}{\sigma\sqrt{\Delta_{t+1}+\dots+\Delta_T}}\left(\ln(\e^{Z_t}/K) + (r+\sigma^2/2)(\Delta_{t+1} +\dots+\Delta_T) \right)$ and $d_2 = d_1-\sigma \sqrt{\Delta_{t+1} +\dots+\Delta_T}$. We thus have exact ground truth benchmark for our estimator $V_{\bm X}$.

After the construction of the estimated value process $V_{\bm X}$, we evaluate $V_{\bm X, t}$ and $V_t$ on the test sample $\bm X_{\mathrm{test}}$, for $t=0,\dots, T-1$. Then we carry out the following four evaluation tasks.

\begin{figure}[p]
    \centering % <-- added
    \begin{subfigure}{0.45\textwidth}
  \includegraphics[width=\linewidth]{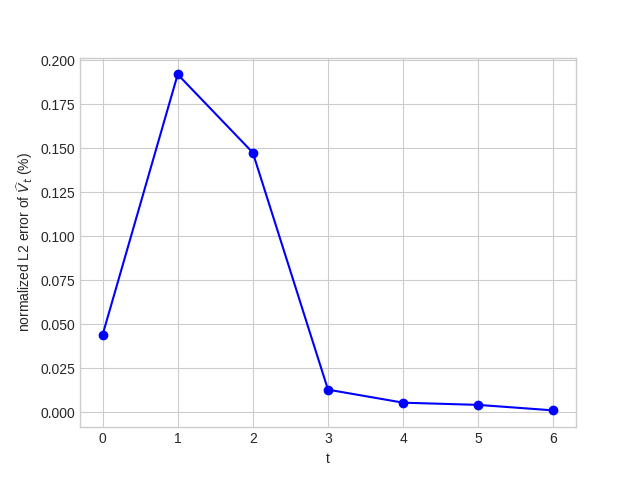}
  \caption{With Random Forest: normalized $L^2_{\Q}$-errors of $V_{\bm X, t}$ as function of $t=0, \dots, T-1$. Values are expressed in \%.}
  \label{error_us_Vt_put_rf}
\end{subfigure}\hfil % <-- added
\begin{subfigure}{0.45\textwidth}
  \includegraphics[width=\linewidth]{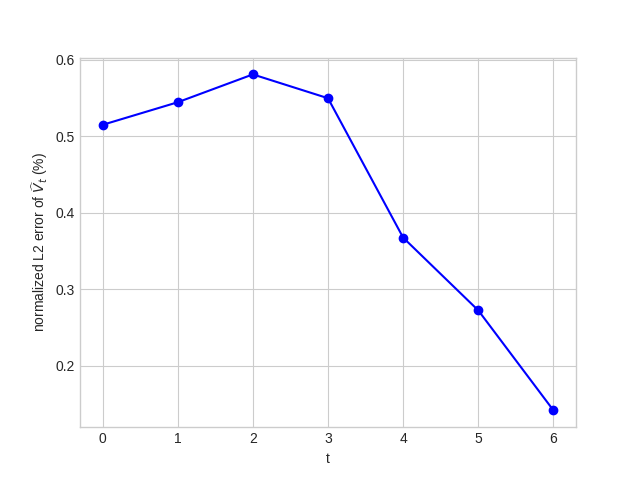}
  \caption{With XGBoost: normalized $L^2_{\Q}$-errors of $V_{\bm X, t}$ as function of $t=0, \dots, T-1$. Values are expressed in \%.}
  \label{error_us_Vt_put_xgb}
\end{subfigure}
\medskip
\begin{subfigure}{0.45\textwidth}
  \includegraphics[width=\linewidth]{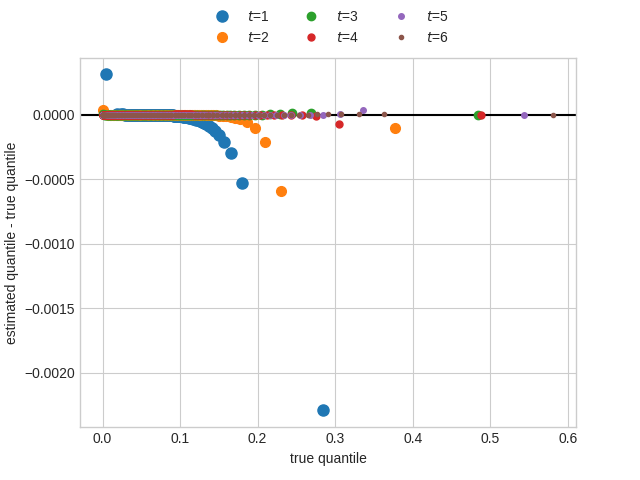}
  \caption{With Random Forest: detrended Q-Q plot of $V_{\bm X, t}$ for $t=1, \dots, T-1$.}
  \label{qqplot_us_Vt_put_rf}
\end{subfigure}\hfil % <-- added
\begin{subfigure}{0.45\textwidth}
  \includegraphics[width=\linewidth]{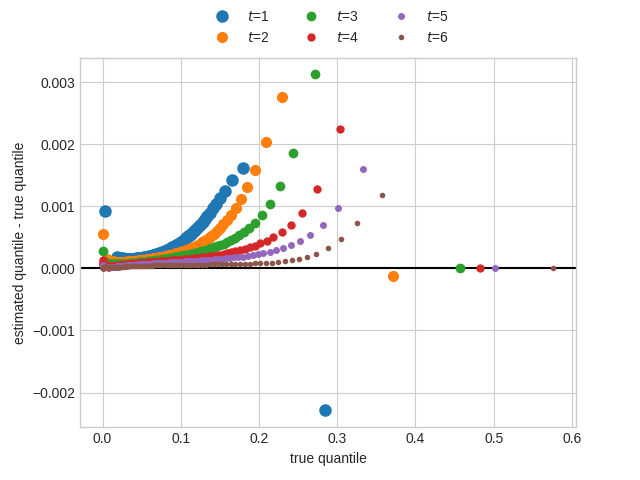}
  \caption{With XGBoost: detrended Q-Q plot of $V_{\bm X, t}$ for $t=1, \dots, T-1$.}
  \label{qqplot_us_Vt_put_xgb}
\end{subfigure}
\medskip
\begin{subfigure}{0.45\textwidth}
  \includegraphics[width=\linewidth]{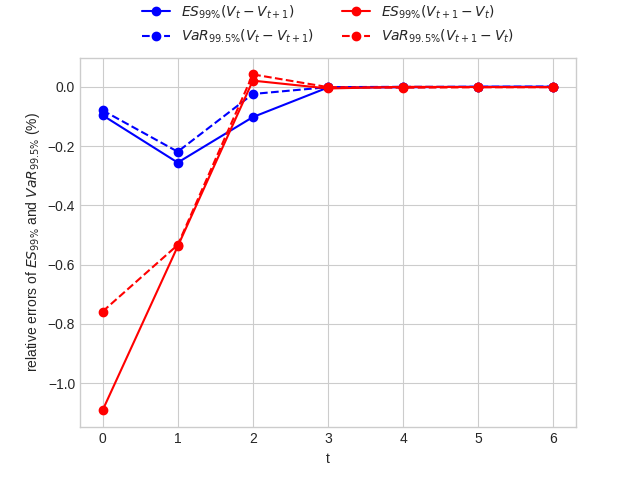}
  \caption{With Random Forest: relative errors of $\mathrm{ES}_{99\%}(V_t-V_{t+1})$, $\mathrm{VaR}_{99.5\%}(V_t-V_{t+1})$, $\mathrm{ES}_{99\%}(V_{t+1}-V_t)$, and $\mathrm{VaR}_{99.5\%}(V_{t+1}-V_t)$ for the estimator $V_{\bm X}$. Values are expressed in \%.}
  \label{normalized_errors_risk_measures_rf}
\end{subfigure}\hfil % <-- added
\begin{subfigure}{0.45\textwidth}
  \includegraphics[width=\linewidth]{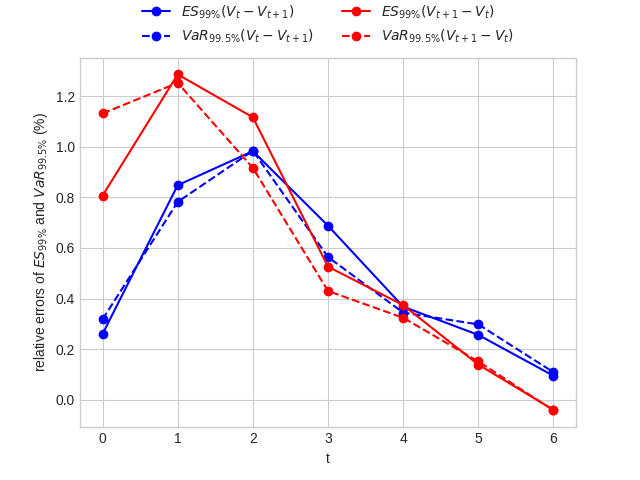}
  \caption{With XGBoost: relative errors of $\mathrm{ES}_{99\%}(V_t-V_{t+1})$, $\mathrm{VaR}_{99.5\%}(V_t-V_{t+1})$, $\mathrm{ES}_{99\%}(V_{t+1}-V_t)$, and $\mathrm{VaR}_{99.5\%}(V_{t+1}-V_t)$ for the estimator $V_{\bm X}$. Values are expressed in \%.}
  \label{normalized_errors_risk_measures_xgb}
\end{subfigure}
\caption{
Results for the Bermudan put with regress-later Random Forest and regress-later XGBoost. The normalized $L^2_\Q$-errors of $V_{\bm X, t}$, $\|V_t-V_{\bm X, t}\|_{2, \Q}/V_0$, for $t=0, \cdots, T-1$, are computed using the test sample and expressed in \%. The detrended Q-Q plots are built using the test sample. They show the detrended quantiles of levels $\{0.001\%, 1\%,2\%,\dots, 100\%\}$ of $V_{\bm X, t}$, for $t=1\cdots, T-1$. The relative errors of value at risk and expected shortfall, computed as $(\text{estimated risk measure minus true risk measure})/\text{true risk measure}$ using the test sample, are expressed in \%.}
\label{fig_us}
\end{figure}

First, we compute the normalized $L^2_\Q$-error $\|V_{\bm X, t}-V_t\|_{2, \Q}/V_0$, for $t=0, \dots, T-1$. Figures \ref{error_us_Vt_put_rf} and \ref{error_us_Vt_put_xgb} show the evolution of the normalized $L^2_\Q$-error of $V_{\bm X, t}$ as function of $t=0, \dots, T-1$. First, we notice that all normalized $L^2_\Q$-errors are below $0.2\%$ and $0.6\%$ with Random Forest and XGBoost, respectively. Second, with the exception of $t=0$, the normalized $L^2_\Q$-errors have a tendency to increase with time to maturity $T-t$. There seems to be an accumulation of errors, due to the estimation of $V_t$ at each induction step $t+1\to t$. Whereas at $t=0$ the errors seem to cancel out across the sample, as $V_{\bm X,0}=C_{\bm X,0}$ is given by the unconditional expectation \eqref{ct_markov}.

Second, we compute the detrended Q-Q plots of $V_{\bm X}$. Figures \ref{qqplot_us_Vt_put_rf} and \ref{qqplot_us_Vt_put_xgb} show detrended Q-Q plots of $V_{\bm X, t}$, for $t=1, \dots, T-1$, using Random Forest and XGBoost, respectively. They are constructed as the detrended Q-Q plots in Section \ref{secexamples}. Specifically, here we draw the detrended Q-Q plots of $V_{\bm X, t}$ using the test sample, for every $t=1, \dots, T-1$. Thereto, for $t\in \{1,\dots, T-1\}$, we compute the empirical left quantiles of $V_{\bm X, t}$ and $V_t$ at levels $\{0.001\%, 1\%, 2\%,\dots,100\%\}$. The detrended quantiles (estimated quantiles minus true quantiles) are then plotted against the true quantiles. We notice that, as function of $t$, the decrease of the normalized $L^2_\Q$-errors of $V_{\bm X, t}$ translates into a flattening of the detrended Q-Q plots. In fact, for illustration, the almost zero normalized $L^2_\Q$-errors of $V_{\bm X, 4}$, $V_{\bm X, 5}$, and $V_{\bm X, 6}$ in Figure \ref{error_us_Vt_put_rf} correspond to almost perfect detrended Q-Q plots of $V_{\bm X, 4}$, $V_{\bm X, 5}$, and $V_{\bm X, 6}$ in Figure \ref{qqplot_us_Vt_put_rf}, i.e., they correspond to detrended Q-Q plots that are almost perfectly aligned with the horizontal black line. Overall, both Random Forest and XGBoost give excellent detrended Q-Q plots of $V_{\bm X, t}$, for every $t=1,\dots, T-1$.

Third, we use our value process estimator $V_{\bm X}$ to compute the value at risk at level $\alpha=99.5\%$, and the expected shortfall at level $\alpha = 99\%$ of $V_{\bm X, t}-V_{\bm X, t+1}$ and $V_{\bm X, t+1}-V_{\bm X, t}$, for every $t=0, \dots, T-1$. $V_{\bm X, t}-V_{\bm X, t+1}$ and $V_{\bm X, t+1}-V_{\bm X, t}$ are the 1-period losses at time $t$ of long position and short position, respectively. We perform the same risk measure computations with our benchmark $V$. Then, we compute the relative errors of risk measures, computed as $(\text{estimated risk measure minus true risk measure})/\text{true risk measure}$ and expressed in \%. Figure \ref{normalized_errors_risk_measures_rf} shows the evolution of relative errors of risk measures $\mathrm{ES}_{99\%}$ and $\mathrm{VaR}_{99.5\%}$ for both long and short positions with Random Forest. Figure \ref{normalized_errors_risk_measures_xgb} shows the same computations with XGBoost. Let's focus on Figure \ref{normalized_errors_risk_measures_rf}. Relative errors of risk measures of long and short positions are all in the intervals $[-0.3\%, 0\%]$ and $[-1.2\%,0.1\%]$, respectively. Furthermore, we highlight that the relative errors for $t\in \{3, 4, 5, 6\}$ are equal to 0\%. The last is in line with the detrended Q-Q plots at $t\in\{3,4, 5, 6\}$ in Figure \ref{qqplot_us_Vt_put_rf}, which are almost perfectly aligned with the horizontal black line.

Fourth, we compute the optimal stopping rule estimator $\tau_{\bm X, 0}^{\star}$ in \eqref{opt_stop_rule_est} for the $n_{\mathrm{test}}$ simulations in $\bm X_{\mathrm{test}}$. Table \ref{table_opt_stop} shows the distribution of $\tau_{\bm X, 0}^{\star}$ using the test sample $\bm X_{\mathrm{test}}$. The distribution of the true optimal stopping rule $\tau_0^\star$ is the Dirac distribution $\delta_{7}(dx)$. We observe that estimation of the distribution of $\tau_0^\star$ is accurate with both Random Forest and XGBoost, and it is XGBoost that outperforms Random Forest. 

%For the latter it is worthwhile to recall Figure \ref{fig_us}, where we see that overall XGBoost outperforms Random Forest in portfolio valuation (see the normalized $L^2_\Q$-errors in Figures \ref{error_us_Vt_put_rf} and \ref{error_us_Vt_put_xgb}), and risk measurements (see the detrended Q-Q plots in Figures \ref{qqplot_us_Vt_put_rf} and \ref{qqplot_us_Vt_put_xgb}, and the normalized errors of risk measures in Figures \ref{normalized_errors_risk_measures_rf} and \ref{normalized_errors_risk_measures_xgb}). In fact, this is an illustration of the no free lunch theorem. This theorem states that an estimator cannot beat all other estimators in all three portfolio valuation, risk measurements, and optimal stopping problems.

\begin{table}
\centering
  \begin{tabular}{|l|l|l|l|l|l|l|l|l|}
\hline
Estimator $|$ time $t$  & 0 &1 &2 &3&4&5&6&7     \\
\hline
XGBoost  & 0& 0.00009& 0.00037& 0.00036& 0.00064& 0.00030& 0.00104& \underline{\textbf{0.99720}}  \\
\hline
Random Forest& 0& 0.00026& 0.00308& 0.00197& 0.00693& 0.00097& 0.00534& \textbf{0.98145}\\
\hline
\end{tabular}
\caption{Distribution of $\tau_{\bm X, 0}^{\star}$, constructed with $V_{\bm X}$ using the test sample $\bm X_{\mathrm{test}}$, using XGBoost and Random Forest. The true distribution of $\tau^\star_0$ is the Dirac distribution $\delta_7(d x)$.}\label{table_opt_stop}
\end{table}

%\newpage
\subsection{Regress-now}
We now compare the above regress-later method for Bermudan options pricing to its regress-now variant, as discussed in Appendix \ref{sec_regnow_reglat} for European style options.

Let us start by introducing the regress-now estimator of the value process of Bermudan options. We shall denote this estimator by $V_{\bm X}^{\mathrm{now}}$. Its construction is as follows. To solve \eqref{eqcondY_US_markov}, proceed backward, where at each time $t=T-1, \dots, 0$ estimate the continuation value function $C_t$ by an ensemble estimator $C_{\bm X, t}^{\mathrm{now}}$. Specifically, assume available a finite i.i.d.\ training sample $\bm{ Z} = (Z^{(1)}, \dots, Z^{(n)})$ drawn from \eqref{eqcondY_US_markov}. And let ${\bm Z}_t = (Z^{(1)}_t, \dots, Z^{(n)}_t )$ denote the $t$-cross-section samples of ${\bm Z}$. Then proceed backward as follows:
\begin{enumerate}
    \item For time $t=T$: set $V_{\bm X, T}^{\mathrm{now}} =  g_T(Z_Z)$.
    \item For any time $t= T-1,\dots,0$: let $C_{\bm X,t}^{\mathrm{now}}$ be an ensemble estimator of $C_{t}$, obtained using the sample ${\bm Z}_{t}$, along with the function values $\bm{V}_{\bm X, t+1}^{\mathrm{now}} = (V_{\bm X, t+1}^{\mathrm{now}}(Z_{t+1}^{(1)}), \dots, V_{\bm X, t+1}^{\mathrm{now}}(Z_{t+1}^{(n)}))$. Then, we set $V_{\bm X, t}^{\mathrm{now}} = \max( g_{t}(Z_{t}), C_{\bm X, t}^{\mathrm{now}})$.
\end{enumerate}
When $V_{\bm X}^{\mathrm{now}}$ is the Random Forest estimator of $V$, we use the RandomForestRegressor class of the library scikit-learn \cite{scikit_learn}, with the hyperparameter values: $M$=500, \textbf{nodesize} = 20, $p=1$, \textbf{sampling
regime}=bootstrapping. When $V_{\bm X}$ is the Gradient Boosting estimator of $V$, we use the XGBRegressor class of XGBoost \cite{che_gue_2016}, with the hyperparameter values: $t=300$, $\textbf{nodesize}=20$, $\textbf{max\_depth}=50$.\footnote{These hyperparameter values give much better numerical results than the default hyperparameter values we used in Section~\ref{sec_us_option_reglater}.}

\begin{figure}[p]
    \centering % <-- added
    \begin{subfigure}{0.45\textwidth}
  \includegraphics[width=\linewidth]{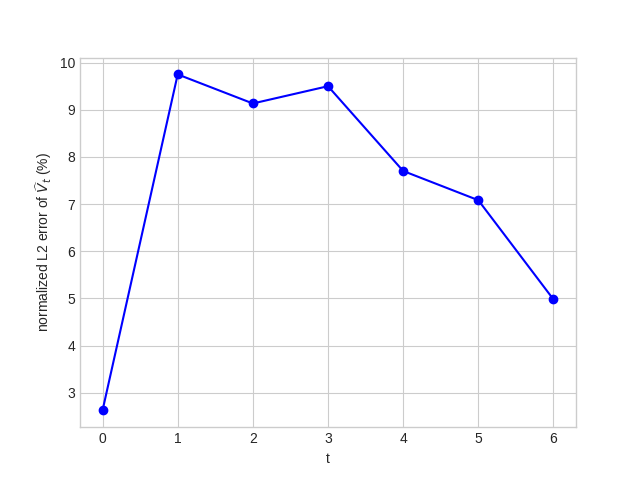}
  \caption{With Random Forest: normalized $L^2_{\Q}$-errors of $V_{\bm X, t}^{\mathrm{now}}$ as function of $t=0, \dots, T-1$. Values are expressed in \%.}
  \label{now_error_us_Vt_put_rf}
\end{subfigure}\hfil % <-- added
\begin{subfigure}{0.45\textwidth}
  \includegraphics[width=\linewidth]{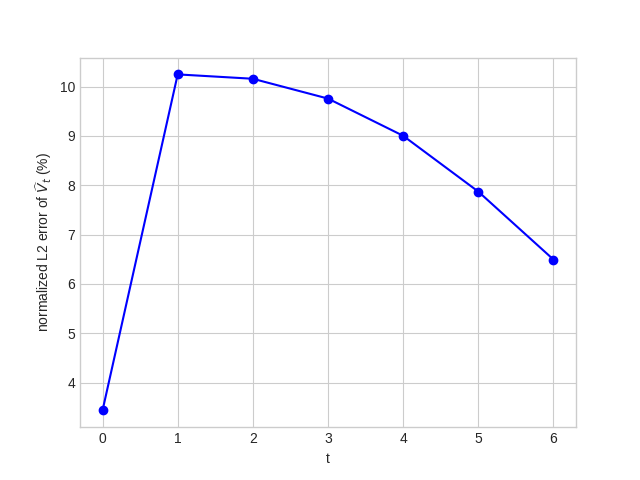}
  \caption{With XGBoost: normalized $L^2_{\Q}$-errors of $V_{\bm X, t}^{\mathrm{now}}$ as function of $t=0, \dots, T-1$. Values are expressed in \%.}
  \label{now_error_us_Vt_put_xgb}
\end{subfigure}
\medskip
\begin{subfigure}{0.45\textwidth}
  \includegraphics[width=\linewidth]{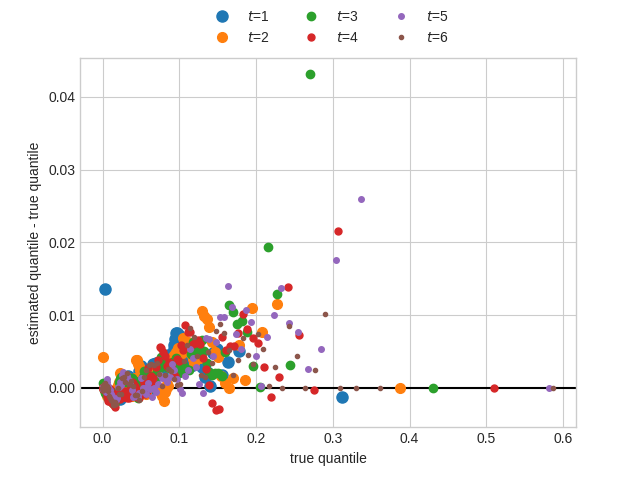}
  \caption{With Random Forest: detrended Q-Q plot of $V_{\bm X, t}^{\mathrm{now}}$ for $t=1, \dots, T-1$.}
  \label{now_qqplot_us_Vt_put_rf}
\end{subfigure}\hfil % <-- added
\begin{subfigure}{0.45\textwidth}
  \includegraphics[width=\linewidth]{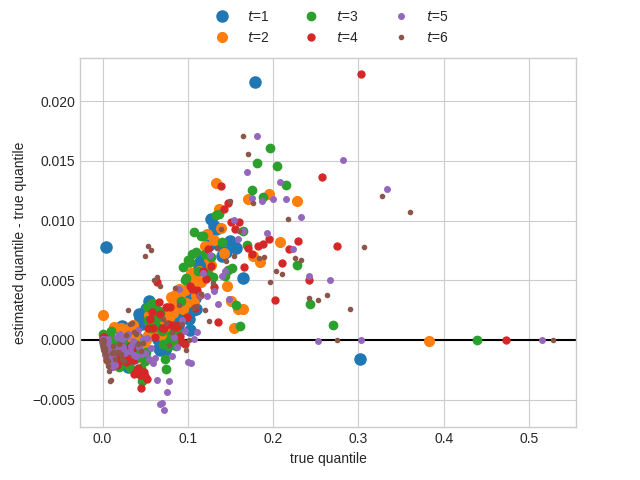}
  \caption{With XGBoost: detrended Q-Q plot of $V_{\bm X, t}^{\mathrm{now}}$ for $t=1, \dots, T-1$.}
  \label{now_qqplot_us_Vt_put_xgb}
\end{subfigure}
\medskip
\begin{subfigure}{0.45\textwidth}
  \includegraphics[width=\linewidth]{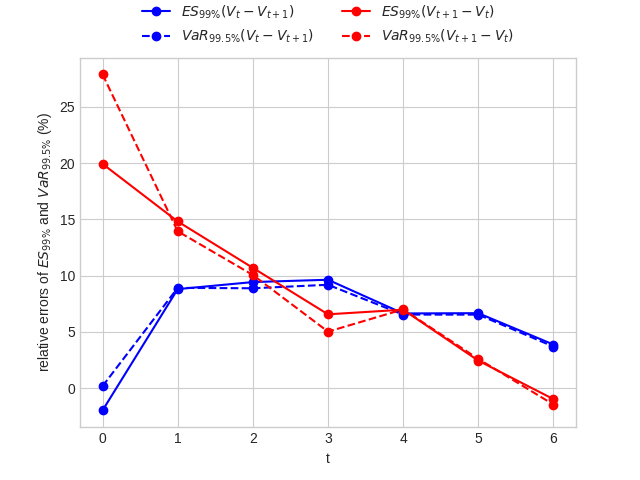}
  \caption{With Random Forest: relative errors of $\mathrm{ES}_{99\%}(V_t-V_{t+1})$, $\mathrm{VaR}_{99.5\%}(V_t-V_{t+1})$, $\mathrm{ES}_{99\%}(V_{t+1}-V_t)$, and $\mathrm{VaR}_{99.5\%}(V_{t+1}-V_t)$ for the estimator $V_{\bm X}^{\mathrm{now}}$. Values are expressed in \%.}
  \label{now_normalized_errors_risk_measures_rf}
\end{subfigure}\hfil % <-- added
\begin{subfigure}{0.45\textwidth}
  \includegraphics[width=\linewidth]{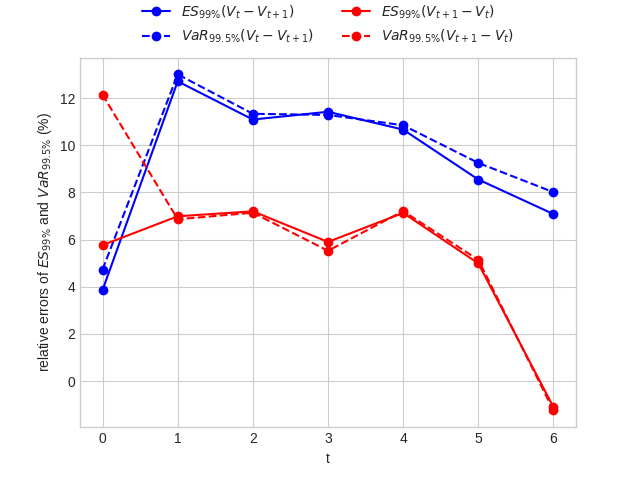}
  \caption{With XGBoost: relative errors of $\mathrm{ES}_{99\%}(V_t-V_{t+1})$, $\mathrm{VaR}_{99.5\%}(V_t-V_{t+1})$, $\mathrm{ES}_{99\%}(V_{t+1}-V_t)$, and $\mathrm{VaR}_{99.5\%}(V_{t+1}-V_t)$ for the estimator $V_{\bm X}^{\mathrm{now}}$. Values are expressed in \%.}
  \label{now_normalized_errors_risk_measures_xgb}
\end{subfigure}
\caption{
Results for the Bermudan put with regress-now Random Forest and regress-now XGBoost. The normalized $L^2_\Q$-errors of $V_{\bm X,t}^{\mathrm{now}}$, $\|V_t-V_{\bm X,t}^{\mathrm{now}}\|_{2, \Q}/V_0$, for $t=0, \cdots, T-1$, are computed using the test sample and expressed in \%. The detrended Q-Q plots are built using the test sample. They show the detrended quantiles of levels $\{0.001\%, 1\%,2\%,\dots, 100\%\}$ of $V_{\bm X,t}^{\mathrm{now}}$, for $t=1\cdots, T-1$. The relative errors of value at risk and expected shortfall, computed as $(\text{estimated risk measure minus true risk measure})/\text{true risk measure}$ using the test sample, are expressed in \%.}
\label{now_fig_us}
\end{figure}

Just as in Section \ref{sec_us_option_reglater} above, we use the test sample $\bm X_{\mathrm{test}}$ to compute the normalized $L^2_\Q$-error $\|V_{\bm X, t}^{\mathrm{now}}-V_t\|_{2, \Q}/V_0$, for $t=0, \dots, T-1$. Figures \ref{now_error_us_Vt_put_rf} and \ref{now_error_us_Vt_put_xgb} show the evolution of the normalized $L^2_\Q$-error of $V_{\bm X, t}^{\mathrm{now}}$ as function of $t=0, \dots, T-1$. First, we notice that all normalized $L^2_\Q$-errors are below $10\%$ and $11\%$ with Random Forest and XGBoost, respectively. Second, with the exception of $t=0$, the normalized $L^2_\Q$-errors have a tendency to increase with time to maturity $T-t$. Again, there seems to be an accumulation of errors, due to the estimation of $C_t$ at each induction step $t+1\to t$. Whereas at $t=0$ the errors seem to cancel out across the sample, as $V_{\bm X,0}=C_{\bm X,0}$ is given by the unconditional expectation \eqref{ct_markov}. Third, by comparing Figures \ref{now_error_us_Vt_put_rf} and \ref{now_error_us_Vt_put_xgb} to Figures  \ref{error_us_Vt_put_rf} and \ref{error_us_Vt_put_xgb}, we highlight the outperformance of our regress-later method over its regress-now variant in terms of normalized $L^2_\Q$-errors.

Then we compute the detrended Q-Q plots of $V_{\bm X}^{\mathrm{now}}$. Figures \ref{now_qqplot_us_Vt_put_rf} and \ref{now_qqplot_us_Vt_put_xgb} show detrended Q-Q plots of $V_{\bm X, t}^{\mathrm{now}}$, for $t=1, \dots, T-1$. They are the counterpart of the detrended Q-Q plots of $V_{\bm X}$ in Figures \ref{qqplot_us_Vt_put_rf} and \ref{qqplot_us_Vt_put_xgb}. By comparing these four figures, we see that the detrended Q-Q plots are of much better quality with regress-later than with regress-now.

Next, we use our value process estimator $V_{\bm X}^{\mathrm{now}}$ to compute the value at risk at level $\alpha=99.5\%$, and expected shortfall at level $\alpha = 99\%$ of $V_{\bm X, t}^{\mathrm{now}}-V_{\bm X, t+1}^{\mathrm{now}}$ and $V_{\bm X, t+1}^{\mathrm{now}}-V_{\bm X, t}^{\mathrm{now}}$, for $t=0, \dots, T-1$. Figures \ref{now_normalized_errors_risk_measures_rf} and \ref{now_normalized_errors_risk_measures_xgb} show relative errors of risk measures with $V_{\bm X}^{\mathrm{now}}$. They are the counterpart of Figures \ref{normalized_errors_risk_measures_rf} and \ref{normalized_errors_risk_measures_xgb}. By comparing these four figures, we see the outperformance of regress-later over regress-now in terms of relative errors of risk measures.

We finish this section by computing the optimal stopping rule estimator $\tau_{\bm X, 0}^{\star, \mathrm{now}}= \inf\{0\le s \le T \mid V_{\bm X, s}^{\mathrm{now}} = g_t(Z_s)\}$ for the $n_{\mathrm{test}}$ simulations in $\bm X_{\mathrm{test}}$. The Table \ref{now_table_opt_stop} is the counterpart of Table \ref{table_opt_stop}. Here again we see the outperformance of regress-later over regress-now in terms of accuracy in the estimation of the optimal stopping rule distribution.

\begin{table}
\centering
  \begin{tabular}{|l|l|l|l|l|l|l|l|l|}
\hline
Estimator $|$ time $t$  & 0 &1 &2 &3&4&5&6&7     \\
\hline
XGBoost & 0& 0.00235& 0.00803& 0.004043& 0.05611& 0.07934& 0.12134& \underline{\textbf{0.69240}} \\
Random Forest& 0& 0.00320& 0.0077& 0.03035& 0.05756& 0.06859& 0.27567& \textbf{0.56386}\\
\hline
\end{tabular}
\caption{Distribution of $\tau_{\bm X, 0}^{\star, \mathrm{now}}$, constructed with $V_{\bm X}^{\mathrm{now}}$ using the test sample $\bm X_{\mathrm{test}}$, using XGBoost and Random Forest. The true distribution of $\tau^\star_0$ is the Dirac distribution $\delta_7(d x)$.
}\label{now_table_opt_stop}
\end{table}

\end{appendix}
%%%%%%%%%%%%%
%%%%%%%%%%%%%%%%%%%%%%%%%%%%%%%%%%%%%%%%%%%%%
%% Bibliography
%%%%%%%%%%%%%%%%%%%%%%%%%%%%%%%%%%%%%%%%%%%%%%%%%%%%%%%%%%
\clearpage
\bibliographystyle{apalike}
\bibliography{Bibliography}

\begin{thebibliography}{}

\bibitem[Archer and Kimes, 2008]{arc_kim_2008}
Archer, K.~J. and Kimes, R.~V. (2008).
\newblock Empirical characterization of random forest variable importance
  measures.
\newblock {\em Comput. Stat. Data Anal.}, 52:2249--2260.

\bibitem[Bartlett et~al., 1998]{sch_etal_1997}
Bartlett, P., Freund, Y., Lee, W.~S., and Schapire, R.~E. (1998).
\newblock {Boosting the margin: a new explanation for the effectiveness of
  voting methods}.
\newblock {\em The Annals of Statistics}, 26(5):1651 -- 1686.

\bibitem[Becker et~al., 2019]{bec_et_al_2019}
Becker, S., Cheridito, P., and Jentzen, A. (2019).
\newblock Deep optimal stopping.
\newblock {\em Journal of Machine Learning Research}, 20(74):1--25.

\bibitem[Biau, 2012]{bia_2012}
Biau, G. (2012).
\newblock Analysis of a random forests model.
\newblock {\em Journal of Machine Learning Research}, 13(38):1063--1095.

\bibitem[Biau and Cadre, 2021]{bia_cad_2017}
Biau, G. and Cadre, B. (2021).
\newblock Optimization by gradient boosting.
\newblock In Daouia, A. and Ruiz-Gazen, A., editors, {\em Advances in
  Contemporary Statistics and Econometrics: Festschrift in Honor of Christine
  Thomas-Agnan}, pages 23--44. Springer International Publishing, Cham.

\bibitem[Biau and Scornet, 2016]{bia_sco_2016}
Biau, G. and Scornet, E. (2016).
\newblock {A random forest guided tour}.
\newblock {\em TEST: An Official Journal of the Spanish Society of Statistics
  and Operations Research}, 25(2):197--227.

\bibitem[Bollerslev, 1986]{boll_1986}
Bollerslev, T. (1986).
\newblock Generalized autoregressive conditional heteroskedasticity.
\newblock {\em Journal of Econometrics}, 31(3):307 -- 327.

\bibitem[Boudabsa and Filipovi\'{c}, 2022]{bou_fil_22}
Boudabsa, L. and Filipovi\'{c}, D. (2022).
\newblock Machine learning with kernels for portfolio valuation and risk
  management.
\newblock {\em Finance Stoch.}, 26(2):131--172.

\bibitem[Breiman, 1996]{bre_1996}
Breiman, L. (1996).
\newblock Bagging predictors.
\newblock {\em Mach. Learn.}, 24(2):123–140.

\bibitem[Breiman, 2001]{bre_2001}
Breiman, L. (2001).
\newblock Random forests.
\newblock In {\em Machine Learning}, pages 5--32.

\bibitem[Breiman et~al., 1984]{bre_et_al_1984}
Breiman, L., Friedman, J., Stone, C., and Olshen, R. (1984).
\newblock {\em Classification and Regression Trees}.
\newblock The Wadsworth and Brooks-Cole statistics-probability series. Taylor
  \& Francis.

\bibitem[Broadie et~al., 2015]{bro_du_moa_15}
Broadie, M., Du, Y., and Moallemi, C.~C. (2015).
\newblock Risk estimation via regression.
\newblock {\em Oper. Res.}, 63(5):1077--1097.

\bibitem[Cambou and Filipovi\'{c}, 2017]{cam_fil_17}
Cambou, M. and Filipovi\'{c}, D. (2017).
\newblock Model uncertainty and scenario aggregation.
\newblock {\em Math. Finance}, 27(2):534--567.

\bibitem[Chen and Guestrin, 2016]{che_gue_2016}
Chen, T. and Guestrin, C. (2016).
\newblock {XGBoost}: A scalable tree boosting system.
\newblock In {\em Proceedings of the 22nd ACM SIGKDD International Conference
  on Knowledge Discovery and Data Mining}, KDD '16, pages 785--794, New York,
  NY, USA. ACM.

\bibitem[Dalcin and Fang, 2021]{mpi4py}
Dalcin, L. and Fang, Y.-L.~L. (2021).
\newblock mpi4py: Status update after 12 years of development.
\newblock {\em Computing in Science Engineering}, 23(4):47--54.

\bibitem[Embrechts, 2009]{emb_2009}
Embrechts, P. (2009).
\newblock Copulas: A personal view.
\newblock {\em The Journal of Risk and Insurance}, 76(3):639--650.

\bibitem[F\"{o}llmer and Schied, 2004]{foe_sch_04}
F\"{o}llmer, H. and Schied, A. (2004).
\newblock {\em Stochastic finance}, volume~27 of {\em De Gruyter Studies in
  Mathematics}.
\newblock Walter de Gruyter \& Co., Berlin, extended edition.
\newblock An introduction in discrete time.

\bibitem[Freund and Schapire, 1996]{fre_sch_1996}
Freund, Y. and Schapire, R.~E. (1996).
\newblock Experiments with a new boosting algorithm.
\newblock In {\em Proceedings of the Thirteenth International Conference on
  International Conference on Machine Learning}, ICML'96, page 148–156, San
  Francisco, CA, USA. Morgan Kaufmann Publishers Inc.

\bibitem[Friedman et~al., 2000]{fri_et_al_2002}
Friedman, J., Hastie, T., and Tibshirani, R. (2000).
\newblock Additive logistic regression: a statistical view of boosting (with
  discussion and a rejoinder by the authors).
\newblock {\em Ann. Statist.}, 28(2):337--407.

\bibitem[Friedman, 2001]{fri_2001}
Friedman, J.~H. (2001).
\newblock Greedy function approximation: A gradient boosting machine.
\newblock {\em Ann. Statist.}, 29(5):1189--1232.

\bibitem[Genuer and Poggi, 2017]{gen_pog_2016}
Genuer, R. and Poggi, J.-M. (2017).
\newblock {Arbres CART et For{\^e}ts al{\'e}atoires,Importance et s{\'e}lection
  de variables}.
\newblock Preprint.

\bibitem[Genuer et~al., 2008]{gen_et_al_2008}
Genuer, R., Poggi, J.-M., and Tuleau, C. (2008).
\newblock Random forests: some methodological insights.
\newblock Preprint.

\bibitem[Genuer et~al., 2010]{gen_et_al_2010}
Genuer, R., Poggi, J.-M., and Tuleau-Malot, C. (2010).
\newblock Variable selection using random forests.
\newblock {\em Pattern Recognition Letters}, 31(14):2225 -- 2236.

\bibitem[Genz, 2000]{genz_2000}
Genz, A. (2000).
\newblock Numerical computation of multivariate normal probabilities.
\newblock {\em Journal of Computational and Graphical Statistics}, 1.

\bibitem[Genz, 2022]{genz_web}
Genz, A. (2022).
\newblock {A}lan {G}enz website, software column.
\newblock \url{http://www.math.wsu.edu/faculty/genz/homepage}.
\newblock Accessed: 2022-03-29.

\bibitem[Genz et~al., 2021]{genz_et_al_mvtnorm}
Genz, A., Bretz, F., Miwa, T., Mi, X., Leisch, F., Scheipl, F., and Hothorn, T.
  (2021).
\newblock {\em {mvtnorm}: Multivariate Normal and t Distributions}.
\newblock R package version 1.1-3.

\bibitem[Glasserman and Yu, 2004]{gla_yu_2004}
Glasserman, P. and Yu, B. (2004).
\newblock Simulation for american options: Regression now or regression later?
\newblock In Niederreiter, H., editor, {\em Monte Carlo and Quasi-Monte Carlo
  Methods 2002}, pages 213--226, Berlin, Heidelberg. Springer Berlin
  Heidelberg.

\bibitem[Gordy and Juneja, 2010]{gor_jun_10}
Gordy, M.~B. and Juneja, S. (2010).
\newblock Nested simulation in portfolio risk measurement.
\newblock {\em Management Science}, 56(10):1833--1848.

\bibitem[Goudenège et~al., 2020]{gou_et_al_2020}
Goudenège, L., Molent, A., and Zanette, A. (2020).
\newblock Machine learning for pricing american options in high-dimensional
  markovian and non-markovian models.
\newblock {\em Quantitative Finance}, 20(4):573--591.

\bibitem[Ke et~al., 2017]{guo_et_al_2017}
Ke, G., Meng, Q., Finley, T., Wang, T., Chen, W., Ma, W., Ye, Q., and Liu,
  T.-Y. (2017).
\newblock Lightgbm: A highly efficient gradient boosting decision tree.
\newblock In Guyon, I., Luxburg, U.~V., Bengio, S., Wallach, H., Fergus, R.,
  Vishwanathan, S., and Garnett, R., editors, {\em Advances in Neural
  Information Processing Systems 30}, pages 3146--3154. Curran Associates, Inc.

\bibitem[Kearns, 1988]{kea_1988}
Kearns, M. (1988).
\newblock Thoughts on hypothesis boosting.
\newblock Unpublished manuscript.

\bibitem[Kearns and Valiant, 1994]{kea_val_1994}
Kearns, M. and Valiant, L. (1994).
\newblock Cryptographic limitations on learning boolean formulae and finite
  automata.
\newblock {\em J. ACM}, 41(1):67–95.

\bibitem[Liaw and Wiener, 2002]{rf_r_pack}
Liaw, A. and Wiener, M. (2002).
\newblock Classification and regression by randomforest.
\newblock {\em R News}, 2(3):18--22.

\bibitem[Loh, 2014]{loh_2014}
Loh, W.-Y. (2014).
\newblock Fifty years of classification and regression trees.
\newblock {\em International Statistical Review}, 82(3):329--348.

\bibitem[Longstaff and Schwartz, 2001]{lon_sch_2001}
Longstaff, F. and Schwartz, E. (2001).
\newblock Valuing american options by simulation: A simple least-squares
  approach.
\newblock {\em Review of Financial Studies}, 14:113--47.

\bibitem[Louppe, 2014]{lou_2014}
Louppe, G. (2014).
\newblock {\em Understanding Random Forests: From Theory to Practice}.
\newblock PhD thesis, University of Li\`ege.

\bibitem[McNeil et~al., 2015]{mcn_et_al_2015}
McNeil, A.~J., Frey, R., and Embrechts, P. (2015).
\newblock {\em Quantitative Risk Management: Concepts, Techniques and Tools}.
\newblock Princeton University Press, USA.

\bibitem[Morgan and Sonquist, 1963]{mor_son_1963}
Morgan, J.~N. and Sonquist, J.~A. (1963).
\newblock Problems in the analysis of survey data, and a proposal.
\newblock {\em Journal of the American Statistical Association},
  58(302):415--434.

\bibitem[Pedregosa et~al., 2011]{scikit_learn}
Pedregosa, F., Varoquaux, G., Gramfort, A., Michel, V., Thirion, B., Grisel,
  O., Blondel, M., Prettenhofer, P., Weiss, R., Dubourg, V., Vanderplas, J.,
  Passos, A., Cournapeau, D., Brucher, M., Perrot, M., and Duchesnay, E.
  (2011).
\newblock Scikit-learn: Machine learning in {P}ython.
\newblock {\em Journal of Machine Learning Research}, 12:2825--2830.

\bibitem[Peskir and Shiryaev, 2006]{pes_shi_2006}
Peskir, G. and Shiryaev, A. (2006).
\newblock {\em Optimal Stopping and Free-Boundary Problems}.
\newblock Basel, Boston: Birkh{\"a}user Verlag.

\bibitem[Prokhorenkova et~al., 2018]{pro_et_al_2018}
Prokhorenkova, L., Gusev, G., Vorobev, A., Dorogush, A.~V., and Gulin, A.
  (2018).
\newblock Catboost: unbiased boosting with categorical features.
\newblock In Bengio, S., Wallach, H., Larochelle, H., Grauman, K.,
  Cesa-Bianchi, N., and Garnett, R., editors, {\em Advances in Neural
  Information Processing Systems 31}, pages 6638--6648. Curran Associates, Inc.

\bibitem[Quinlan, 1993]{qui_1993}
Quinlan, J.~R. (1993).
\newblock {\em C4.5: Programs for Machine Learning}.
\newblock Morgan Kaufmann Publishers Inc., San Francisco, CA, USA.

\bibitem[Revuz and Yor, 1994]{rev_yor_94}
Revuz, D. and Yor, M. (1994).
\newblock {\em Continuous martingales and {B}rownian motion}, volume 293 of
  {\em Grundlehren der Mathematischen Wissenschaften [Fundamental Principles of
  Mathematical Sciences]}.
\newblock Springer-Verlag, Berlin, second edition.

\bibitem[Schapire, 1990]{Sch_1990}
Schapire, R.~E. (1990).
\newblock The strength of weak learnability.
\newblock In {\em Machine Learning}.

\bibitem[Scornet, 2016]{sco_2016}
Scornet, E. (2016).
\newblock On the asymptotics of random forests.
\newblock {\em Journal of Multivariate Analysis}, 146:72 -- 83.
\newblock Special Issue on Statistical Models and Methods for High or Infinite
  Dimensional Spaces.

\bibitem[Scornet et~al., 2015]{sco_et_al_2015}
Scornet, E., Biau, G., and Vert, J.-P. (2015).
\newblock Consistency of random forests.
\newblock {\em Ann. Statist.}, 43(4):1716--1741.

\bibitem[Sklar, 1959]{skl_1959}
Sklar, M. (1959).
\newblock Fonctions de r\'{e}partition \`a {$n$} dimensions et leurs marges.
\newblock {\em Publ. Inst. Statist. Univ. Paris}, 8:229--231.

\bibitem[Team, 2022]{joblib}
Team, J.~D. (2022).
\newblock Joblib: running python functions as pipeline jobs.
\newblock \url{https://joblib.readthedocs.io/en/latest/}.
\newblock Accessed: 2022-03-29.

\bibitem[{Tsitsiklis} and {van Roy}, 1999]{tsi_roy_1999}
{Tsitsiklis}, J.~N. and {van Roy}, B. (1999).
\newblock Optimal stopping of markov processes: Hilbert space theory,
  approximation algorithms, and an application to pricing high-dimensional
  financial derivatives.
\newblock {\em IEEE Transactions on Automatic Control}, 44(10):1840--1851.

\bibitem[{Virtanen} et~al., 2020]{scipy}
{Virtanen}, P., {Gommers}, R., {Oliphant}, T.~E., {Haberland}, M., {Reddy}, T.,
  {Cournapeau}, D., {Burovski}, E., {Peterson}, P., {Weckesser}, W., {Bright},
  J., {van der Walt}, S.~J., {Brett}, M., {Wilson}, J., {Jarrod Millman}, K.,
  {Mayorov}, N., {Nelson}, A. R.~J., {Jones}, E., {Kern}, R., {Larson}, E.,
  {Carey}, C., {Polat}, {\.I}., {Feng}, Y., {Moore}, E.~W., {Vand erPlas}, J.,
  {Laxalde}, D., {Perktold}, J., {Cimrman}, R., {Henriksen}, I., {Quintero},
  E.~A., {Harris}, C.~R., {Archibald}, A.~M., {Ribeiro}, A.~H., {Pedregosa},
  F., {van Mulbregt}, P., and {Contributors}, S. .~. (2020).
\newblock {SciPy 1.0: Fundamental Algorithms for Scientific Computing in
  Python}.
\newblock {\em Nature Methods}, 17:261--272.

\end{thebibliography}
\end{document}